\documentclass[useAMS,usenatbib]{mn2e}

\usepackage{graphicx}
\usepackage{lscape}
\usepackage{rotating}
\usepackage{pst-all}
\usepackage{amssymb}
\newcommand{\etal}{{\it et~al.}}
\newcommand{\mstar}{M$_\odot$}
\newcommand{\microm}{$\mu$m}
\newcommand{\IRAS}{{\it IRAS}}
\newcommand{\MSX}{{\it MSX}}
\newcommand{\Spitzer}{{\it Spitzer}}
\newcommand{\uchii}{{UC H{\scriptsize II}}}
\newcommand{\hii}{{H{\scriptsize II}}}
\newcommand{\lum}{L$_\odot$} 
\newcommand{\rsun}{R$_\odot$} 

\newcommand{\dd}{\mathrm{d}}

\newcommand{\rmchr}[1]{}

\title[SED modelling of Southern candidate massive protostars]{Spectral energy distribution modelling of Southern candidate massive protostars using the Bayesian inference method.}
\author[T. Hill \etal]{T. Hill$^{1,2,3,}$\thanks{E-mail:thill@astro.ex.ac.uk},  C. Pinte $^1$, V. Minier$^{4,5}$, M. G. Burton$^3$ and M. R. Cunningham$^3$\\
$^{1}$ School of Physics, University of Exeter, Stocker Road, EX4 4QL, Exeter, UK\\
$^{2}$ Leiden Observatory, Leiden University, PO BOX 9513, 2300 RA Leiden, the Netherlands\\ 
$^{3}$ School of Physics, University of New South Wales, Sydney 2052, NSW, Australia\\
$^4$ CEA, DSM, IRFU, Service d'Astrophysique, 91191 Gif-sur-Yvette, France\\
$^5$ Laboratoire AIM, CEA/DSM - CNRS - Universit\'e Paris Diderot, IRFU/Service d'Astrophysique, CEA-Saclay, 91191 \\Gif-sur-Yvette, France 
}

\begin{document}

\date{Accepted/Received}

\pagerange{\pageref{firstpage}--\pageref{lastpage}} \pubyear{2008}

\maketitle

\label{firstpage}

\begin{abstract}

  Concatenating data from the millimetre regime to the infrared, we have performed spectral energy distribution modelling for 227 of the 405 millimetre continuum sources of \citet{hill05} which are thought to contain young massive stars in the earliest stages of their formation. Three main parameters are extracted from the fits: temperature, mass and luminosity. The method employed was Bayesian inference, which allows a statistically probable range of suitable values for each parameter to be drawn for each individual protostellar candidate. This is the first application of this method to massive star formation.

The cumulative distribution plots of the SED modelled parameters in this work indicate that collectively, the sources without methanol maser and/or radio continuum associations (MM-only cores) display similar characteristics to those of high mass star formation regions. Attributing significance to the marginal distinctions between the MM-only cores and the high-mass star formation sample  we draw hypotheses regarding the nature of the MM-only cores, including the possibility that the population itself is comprised of different types of source, and discuss their role in the formation scenarios of massive star formation. In addition, we discuss the usefulness and limitations of SED modelling and its application to the field. From this work, it is clear that within the valid parameter ranges, SEDs utilising current far-infrared data can not be used to determine the evolution of massive protostars or massive young stellar objects. 

\end{abstract}


\section{Introduction}\label{sec:intro}

  The study of massive stars and their natal molecular environments has seen a surge in interest and study in recent years. Despite this, precise scenarios for the formation of massive stars, especially of the earliest stages of their evolution, are still not forthcoming. This may be attributed to a number of factors. The very nature of massive star formation (i.e. rare, rapid, clustered and distant) impedes study of the earliest stages of their evolution, which are not easily distinguished. Massive stars also form in turbulent and evolving environments, e.g. jets, disks and outflows, which contribute to hinder extraction and interpretation of information from these regions. Additionally, instrumental limitations, specifically the resolution of current instruments, are not sufficient to probe the inner most workings of the cocoons in which massive stars are forming.

   Previous studies have identified associations between young massive stars and methanol masers  \citep{pestalozzi05}, Ultra Compact \hii\, (\uchii) regions \citep{thompson06}, \IRAS\, colour selected sources \citep{wood89morph} as well as \MSX\, colour selected sources \citep{lumsden02}. Methanol masers and \uchii\, regions, in particular, are thought to feature prominently in the earliest stages of massive star formation \citep[c.f.][and references within]{batrla87, caswell95, minier01, beuther02, faundez04, williams04, thompson06}.

  Early work \citep[e.g.][]{walsh98} suggested that the methanol maser was the earliest indicator of massive star formation prior to the onset of \hii\, regions that are signposted by radio continuum emission. More recent work has focused on finding a precursor stage to that signposted by radio continuum emission, which would mark the {\it very} earliest stages of massive star formation. The hot molecular core (HMC) and infrared dark clouds are such objects proposed to satisfy this criterion \citep[cf.][]{olmi96, osorio99, hill05, rathborne07}. The methanol masers likely form during these stages \citep{hill05,longmore06, longmore07}

   \citet[hereafter Paper I]{hill05} undertook a SIMBA millimetre continuum emission survey toward regions displaying evidence of massive star formation, in search of cold cores that would mark the earliest stages of their evolution. This survey revealed each of the methanol maser and radio continuum sources targeted to be associated with millimetre continuum emission. Interestingly, this survey also revealed evidence of star formation clearly offset from, and devoid of, both the methanol maser and radio continuum sources targeted. These sources were dubbed `MM-only' cores\footnote{In this paper, the term `cores' refers to molecular cloud fragments that were detected through millimetre dust continuum emission. These `cores' have sizes and masses that span a large range which qualifies them to form many young stellar objects or even protostellar clusters. Although they are more widely labelled as `clumps' in the literature, for consistency with our earlier work we preferentially use the term `cores' in this work.}.

Preliminary analysis showed these MM-only sources to be smaller and less massive than cores harbouring a methanol maser and/or an \uchii\, region. This conclusion was drawn however assuming a constant temperature of 20\,K across the sources in the sample. At least 45 per cent of these `MM-only' sources are also without mid-infrared \MSX\, emission.  It was consequently proposed that the MM-only core is a possible precursor to the methanol maser stage of massive star formation, and thus traces an even earlier stage - perhaps even {\em the} earliest stage in the formation of massive stars. Follow-up submillimetre observations of these MM-only cores  \citep[][hereafter Paper II]{hill06} revealed each of them to be associated with submillimetre continuum emission confirming their association with cold, deeply embedded objects. 

  In order to ascertain the nature of the MM-only cores and any role that they play in the formation and evolution of massive stars, it is necessary to determine their ambient physical conditions, such as temperature, luminosity and mass. Only in light of this information is it possible to characterize the MM-only core and address hypotheses regarding their formation and/or whether they are indicative of the earliest stages of massive star formation. Additionally, in order to put the MM-only core into context within an evolutionary sequence for massive star formation, a wide cross-section of sources suspected of being at various evolutionary stages is required.

   In this paper we combine the (sub)millimetre data from our earlier work (Papers I and II), with existing submillimetre data \citep{p-p00, walsh03, thompson06}, together with archival \MSX\, data where applicable, and \IRAS\, data (more often than not as upper limits) to draw spectral energy distribution (SED) diagrams for a large sample of sources detected with SIMBA. The sample itself is a cross-section of sources suspected of representing different evolutionary stages of massive star formation (see $\S$\,\ref{sec:sample}).

\section{Multiwavelength Compilation of the SIMBA Sources}\label{sec:data}
   
   In this section, we outline the source selection criteria, explore each of the wavebands used for the SED fitting, and explore infrared associations for the sample.

\subsection{The sample origin: (sub)millimetre observations}\label{sec:sample}

Our 1.2\,mm SIMBA survey (Paper I) revealed a total of 405 millimetre continuum sources, a large number of which are MM-only cores as introduced in section \ref{sec:intro}. 

The SIMBA sample is comprised of four distinct classes of source, distinguished by the presence, or lack thereof, of methanol maser and radio continuum tracers. Class M sources are millimetre sources with methanol maser sites but are devoid of radio continuum emission. Class R sources are millimetre sources with radio continuum emission but without methanol maser emission. Class MR sources are millimetre sources with both methanol maser and radio continuum emission. The fourth class of source is the MM-only sample which is comprised of sources with millimetre continuum emission, but without methanol maser sites or \uchii\, regions.

  The observation and data reduction method of each of the SIMBA and SCUBA surveys, are as described in Papers I and II, respectively and the reader is referred to these for more information. 

   The millimetre (1.2\,mm) and submillimetre (450 \& 850\microm) fluxes used in this work were extracted directly from the (sub)millimetre continuum maps using the respective data reduction and analysis packages (see Papers I and II). 
 This procedure involved distinguishing the source from the background and subtracting the latter from the former using apertures defined to 10 per cent contour level (of the peak flux) for each of the SIMBA and SCUBA sources, respectively.

The definition of the source size (contour level of 10 per cent) is critical for SED analysis as it influences the amount of flux input into the SED as well as the resultant parameters from the model. Comparison of the integrated flux determined for a sample of sources to a contour level of 5, 10 and 20 per cent of the peak flux of the source, reveals an integrated flux difference of less than 10 per cent for a 5 per cent contour and less than 15 per cent for a 20 per cent contour, when compared with our assumed size of a 10 per cent contour. As discussed in Section \ref{sec:model}, we assume a 20 per cent flux error for SED analysis which accounts for any ambiguity in source size.

\subsection{Infrared data}

\subsubsection{Infrared Astronomical Satellite (\IRAS)}\label{IRAS}

    The Infrared Astronomical Satellite (\IRAS) was a joint scientific project between the United Kingdom, the Netherlands and the United States. \IRAS\,' mission was to perform a sensitive and unbiased all-sky survey centred at four wavebands in the infrared regime: 12, 25, 60, 100\,\microm. \IRAS\, was launched in January 1983 and ended its mission ten months later in November, after surveying 96 per cent of the sky\footnote{See http://irsa.ipac.caltech.edu/IRASdocs/iras.html}. The angular resolution of \IRAS\, varies between about 30 arcsec at 12\microm\, to about 2 arcmin at 100\microm. For more detail about the design and performance of \IRAS\, refer to \citet{beichman88}.

   The \IRAS\, Sky Survey Atlas (ISSA) is a publicly available set of FITS images of the infrared sky, while the \IRAS\, point source catalogue (PSC) provides flux estimates of sources detected by \IRAS\, for each of the wavebands of observation.

   The \IRAS\, infrared fluxes at 60 and 100\, \microm\, have been extracted from the \IRAS\, PSC for those sources with an \IRAS\, association. They are also used for the purposes of upper limit constraints to spectral energy distribution analysis in the absence of a direct association. Note that we do not use the other two \IRAS\, wavebands at 12 and 25\,\microm\, as they overlap that of the \MSX\, satellite, which achieved better angular resolution (see following section \ref{sec:MSX}).
 
\subsubsection{Midcourse Space Experiment (\MSX)}\label{sec:MSX}

   The Midcourse Space Experiment (\MSX) was a multi-discipline experiment sponsored by the Ballistic Missile Defence Organisation. Launched April 24, 1996 this infrared satellite operated at a temperature of 11 to 12\,K, and spanned the infrared regime from 4.2\,-\,26\,\microm. The four main wavebands of \MSX\, are centred at 8.3, 12.1, 14.7 and 21.3\,\microm.

   The \MSX\, Galactic Plane Survey mapped the Galactic Plane for $|b|<$\,5$^\circ$, and surveyed the part of the sky missed by \IRAS\, in the $``$Survey of Areas Missed by \IRAS'', as well as other surveys. An overview of the astronomical experiments conducted with \MSX\, is given by \citet{price95}, while a complete description of the experiments and data processing is given in \citet{price01}. 
   
   The infrared instrument on \MSX\, SPIRIT III, had a spatial resolution of 18.3 arcsec, and a sensitivity of 0.1\,Jy at 8.3\microm. 

   The \MSX\, images were examined for sources appearing in the SIMBA source list. For those sources with a direct \MSX\, association, the mid-infrared flux density was determined from the calibrated images. The maps were converted from B1950 Galactic coordinates to J2000 equatorial coordinates. The maps were then converted from  W\,m$^2$\,sr$^{-1}$  to Jy, allowing for 6 arcsec square pixels, and an additional factor of 1.133 to convert from square pixels into a gaussian area. The final conversion factors for each of the wavebands were 6.84$\times 10^3$, 2.74$\times10^4$, 3.08$\times 10^4$ and 2.37$\times 10^4$ Jy per W\,m$^{2}$sr$^{-1}$ at 8.3, 12.1, 14.7 and 21.3\,\microm, respectively.

The flux of the source was then measured using the {\sc karma/kvis}\footnote{http://www.atnf.csiro.au/computing/software/karma/} package by applying an aperture around the sources and at various points in the image considered to be the background. Contour levels of 10 per cent of the peak source flux were overlaid and used to define the `source' aperture size. The resultant flux of the source was then determined by measuring the flux inside each of the source and background apertures and subtracting the latter from the former. In most instances, the fluxes which were determined from the \MSX\, images were consistent with (to within 10 per cent of) the flux reported by the \MSX\, catalogue. However, as there are known problems with the fluxes reported in the \MSX\, catalogue, we cautiously opted to manually determine fluxes from the \MSX\, images for all sources.

   For those sources with an \MSX\, association, the infrared \MSX\, emission is used to draw the SED as described in section \ref{sec:sedfits}. 

\subsection{Correlating the SIMBA data with other data}\label{sec:corr} 

  The aim of this paper is to draw spectral energy distribution (SED) diagrams for individual sources in the SIMBA source list of Paper I. Submillimetre associations with these data were explored in Paper II. For those SIMBA cores which are resolved by SCUBA into multiple components, the corresponding submillimetre flux for the SIMBA sources has been used to draw the SED, with individual submillimetre components summed together.

   Extracting source specific information from all sky surveys can be difficult due to confusion in the images. The infrared images from the \IRAS\, and \MSX\, satellites are examined here with respect to each of the sources in the SIMBA images.

   As a consequence of the poor spatial resolution of \IRAS\, compared to that of SIMBA (24 arcsec), it was often not possible to conclude an \IRAS\, association with an individual SIMBA source.  Generally the entire SIMBA map  (240 $\times$ 480 arcsec$^2$) falls within a single \IRAS\, source, that is, the SIMBA instrument resolved the corresponding \IRAS\, source into multiple millimetre components. Very few sources in the sample had a direct correlation with the \IRAS\, peak of emission and hence an \IRAS\, source. An \IRAS\, flux was generally only used for isolated SIMBA sources. For all other sources, the IRAS flux corresponding to the nearest coincident methanol maser and/or radio continuum source was used as an upper limit in the SEDs. In both of these instances, the IRAS flux was taken directly from the PSC. In the few cases where a SIMBA source was completely devoid of \IRAS\, emission, falling in diffuse background emission instead, an upper limit was obtained from the \IRAS\, images according to the procedure described for the \MSX\, data (section \ref{sec:MSX}), and is included on the SED plots.

Considering the resolution of \IRAS\, the usefulness of these data for characterisation and constraint of the cold component of our SED fits is questionable. It may be argued instead that \Spitzer\, data would be a better choice, especially the 70\,\microm\, data which easily supersedes the resolution of the 60\,\microm\, \IRAS\, data. We have however, elected not to use the \Spitzer\, data for our SED fits primarily as these data are an, as yet, unpublished data set. While the data are available for download, it contains many artefacts and saturated values, which we are unable to quantify.

  As mentioned in section \ref{sec:MSX}, the \MSX\, satellite produced higher resolution images than that of \IRAS. However as a result of an excess of mid-infrared emission in the fields examined,  it is often not possible to distinguish individual associations due to confusion. For instance, there can be extended PAH emission in the \MSX\, 8, 12 and 14\,\microm\, maps, which makes it difficult to extract the flux associated with dust emission from a core. Consequently, although a SIMBA source may not have a {\em direct} association with a \MSX\, source, it does not eliminate that same source from having associated mid-infrared emission. Often, a SIMBA source falls within diffuse mid-infrared \MSX\, emission, yet they are not directly associated with a \MSX\, peak of emission (i.e. a \MSX\, source). For any SIMBA source where a \MSX\, association was ambiguous, \MSX\, data were not used. \MSX\, emission associated with multiple SIMBA sources was not used unless it was obvious which SIMBA source is dominated by the \MSX\, flux.

   In the case of G\,49.49-0.37, the SCUBA images reveal it to  be quite complicated at both 450 and 850\microm, resolving the SIMBA sources into multiple components. The SIMBA sources also are only partially sampled by our SCUBA data. Due to this inadequate sampling, as well as confusion for associations with \MSX\, and \IRAS, SED analysis was not performed for this entire region (20 sources).


\section{Spectral Energy Distribution Analysis}\label{sec:sedfits}

   The concatenation of the (sub)millimetre and infrared data described in Section \ref{sec:data} enables the spectral energy distribution of each of the SIMBA sources to be drawn.

\subsection{Modelling procedure}\label{sec:model}

We have modelled our sample according to a simple two-component model denoting a central warm core surrounded by a colder dust envelope \citep[see][]{minier05}. The `hot' component of this model is assumed to radiate as a blackbody sphere, whilst the `cold' component accounts for optically thin emission from the dust. The emerging spectrum is then defined by:

\begin{equation}\label{eq:sed_fit}
F_\nu = \left[\pi B_\nu(T_{\mathrm{hot}})\, R_{\mathrm{hot}}^2 +
B_\nu(T_\mathrm{{cold}})\, M_\mathrm{cold}\, \kappa(\nu) \right] / d^2
\end{equation}

\noindent
where $F_\nu$ is the flux density of the source, R$_\mathrm{hot}$ is the radius of the hot component of the source, $B_\nu$ is the Planck function for a temperature of $T_\mathrm{hot}$ and $T_\mathrm{cold}$, $M_\mathrm{cold}$ is the mass of the cold component, $d$ the distance to the source and $\kappa(\nu)$ is the  mass absorption coefficient. In this instance, $\kappa(\nu)$ is assumed to vary as a power-law with $\kappa(\lambda) = \kappa_0\ (\lambda /\lambda_0)^{-2}$\,cm$^2$ per gram of dust at $\lambda_0 = 1.2$\,mm, as per the opacity models of \citet{ossenkopf94} (c.f. \citealp{minier05}) where $\kappa_0 = 1.0$\,cm$^2$ g$^{-1}$. We assume a dust to gas ratio of 100.
 
Equation~\ref{eq:sed_fit} solves for four parameters: $R_\mathrm{hot}$, $T_\mathrm{hot}$, $T_\mathrm{cold}$ and $M_\mathrm{cold}$. Due to the ambiguities with SED fitting, it is not possible to constrain each of these four parameters independently. The robust estimation for the range of validity of the parameters instead requires the potential correlations between each of the parameters to be taken into consideration. With this in mind, we systematically explored a grid of models by varying all four free parameters in order to ascertain a solution. The range of values explored with the fitting procedure are summarized in Table~\ref{tab:param}. Spectral energy distributions were calculated for each combination of these parameters, resulting in a total of 6.25 million synthetic SEDs.

\begin{table}
  \centering
 \caption{Range of values explored for each of the parameters of interest from the fitting procedure using log-space sampling.}
  \begin{tabular}{lcc}
    \hline
    Parameter & range & no. of values sampled \\
    \hline
    $T_\mathrm{cold}$ & 2.73 \ldots 100\,K & 100\\
    $M_\mathrm{cold}$ & 1 \ldots 10$^6$\,\mstar\, & 100 \\
    $T_\mathrm{hot}$ &  100 \ldots 1\,500\,K & 25\\
    $R_\mathrm{hot}$ &  $10^3$ \ldots $10^6$\,\rsun\, & 25 \\
    \hline
  \end{tabular}
 
  \label{tab:param}
\end{table}

Comparisons between the models and the observations were drawn according to a reduced $\chi^2$ value. For each of the individual sources, we computed a table of reduced $\chi^2$ sampling of the whole parameter space. That is, the observed SEDs were compared to all synthetic SEDs according to the calculated  reduced $\chi^2$ values. These tables of reduced $\chi^2$ values are used in the following section to estimate the range of validity for each of the different parameters via the Bayesian inference method.

For the purpose of the $\chi^2$ calculations, each observational flux was assumed to have a flux error of 20 per cent, except for those 450\,\microm\, submillimetre and \IRAS\, data where a 40 per cent error was used. The error estimates for the (sub)millimetre data are consistent with Paper II, whilst for the infrared data these estimates are consistent with \citet{minier05} and deemed reasonable for fitting purposes. 

For those sources devoid of coincident mid-infrared \MSX\, emission, we have fit the (sub)millimetre data for the cold component of Equation~\ref{eq:sed_fit} only, similar to what was done by \citet{minier05}. In the absence of an \IRAS\, association we have used the \IRAS\, flux as an upper limit in the fitting procedure. 

The free-free emission contribution to the millimetre fluxes of the sources in our sample is expected to be minor. Comparison of cm-band radio continuum fluxes with our 1.2\,mm continuum fluxes indicates that the free-free contamination is typically a few percent, even for bright sources such as G\,5.89\,-\,0.39, which has a free-free contamination to the 1.2\,mm flux of not more than 5 per cent. We therefore do not expect free-free emission to significantly influence our fits and do not consider it in the fitting procedure.

The luminosity of each source was determined through integration of Equation~\ref{eq:sed_fit}. We opt to limit the range for luminosity integration between 1.2\,mm and 8\,\microm, corresponding to the region encompassed by our data. At longer wavelengths, there are no data to constrain the SED curve, the shape of which is then reliant upon the model alone. For consistency, we thus also limit the shorter wavelengths to the actual data. We stress that the resultant luminosity is representative of the luminosity between the range integrated (1.2\,mm and 8\,\microm) only and is not a direct representation of the bolometric luminosity of the source (i.e. it is a lower limit to the luminosity).
In order to ascertain the effect that limiting the integration range had on the resultant luminosity, we also integrated the luminosity over the wavelength range 0.1\,\microm\, to 3\,mm. Although the luminosity for this range was roughly only 5 per cent greater on average than the range encompassing our data (1.2\,mm to 8\,\microm), individual sources could vary as much as 40 per cent. Despite the differences in the luminosity determined for each range, varying the integration range had negligible effect on the shape of the cumulative distributions in Fig. \ref{fig:cumul}.

\subsection{Validity range of parameters}

To determine the range of validity for each of the four free parameters, or two free parameters for the single component analysis, we used a Bayesian inference method \citep{press92,lay97,pinte07,pinte08}. This technique allows us to estimate the probability of occurrence of each parameter value. The relative probability of a single point of the parameter space  (\emph{i.e.} one model) is proportional to $\exp (-\chi^2/2)$, where $\chi^2$ refers to the reduced $\chi^2$ of the corresponding model. All probabilities are normalized at the end of the procedure so that the sum of the probabilities of all models over the entire grid is equal to 1. 

The Bayesian inference method of SED modelling does not only give {\it the} best value for each parameter, it also produces a range of suitable values for each parameter, with each value having an associated probability of occurrence.

The Bayesian method relies on \emph{a priori} probabilities for the parameters. For the purposes of our analysis, we assume that we do not have any \emph{preliminary available information}, choosing instead a uniform a priori probability, which corresponds to a logarithmic sampling of the parameters. 

\begin{figure*}
  \includegraphics[angle=270,width=\hsize]{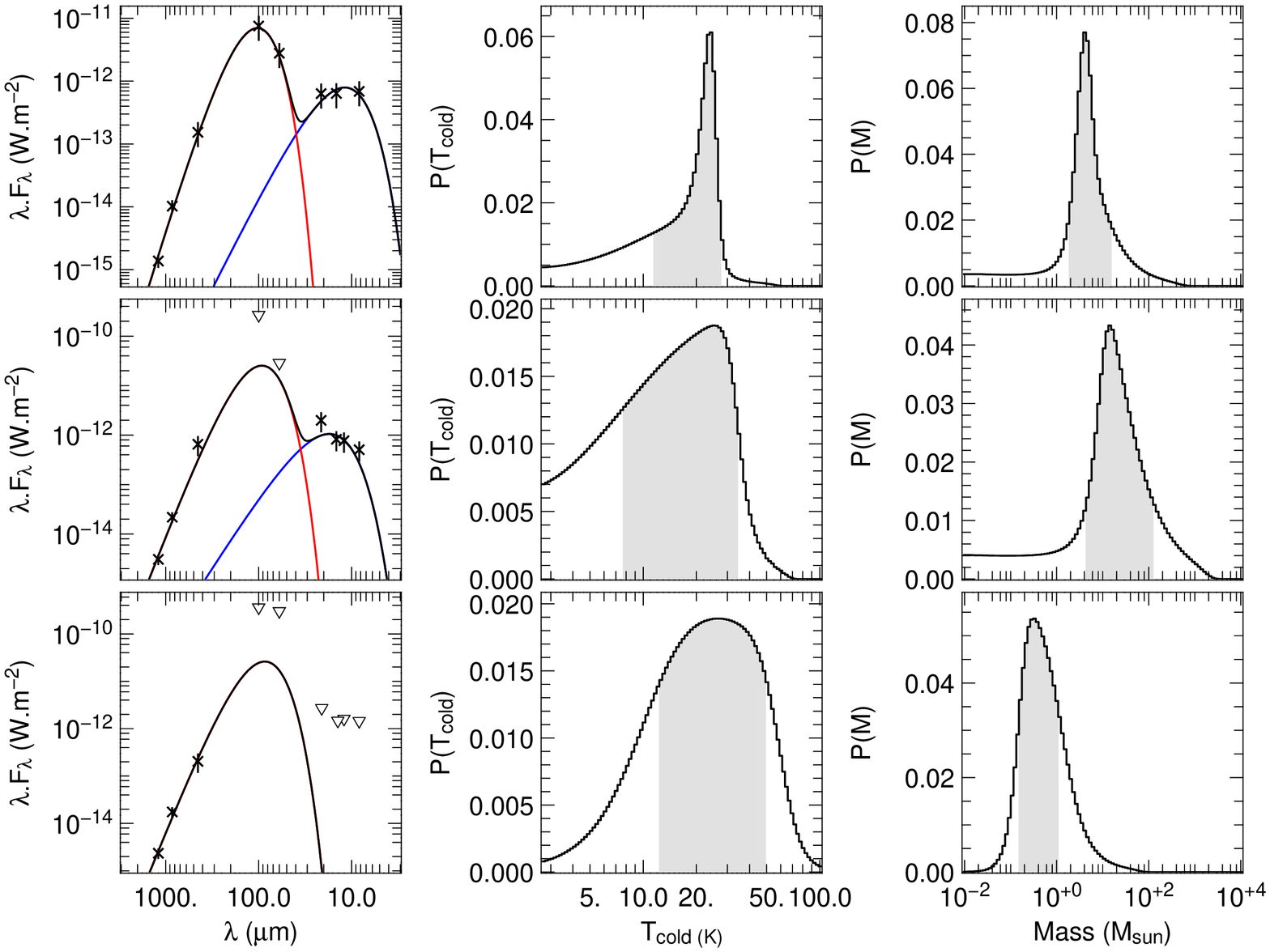}
\caption{SED fits and probability distributions of the temperature and the mass for three sources, from top to bottom: G\,12.02\,-\,0.03, G\,0.21\,-\,0.00 and G\,10.288\,-\,0.127. {\it Left:} example spectral energy distributions. The blue and red lines represent the hot and cold components, respectively. The black line represents the total SED fit. Crosses are observational data points whilst triangles are upper limits. {\it Centre} and {\it right:} histograms of marginal probability distributions for the temperature and mass of the cold component, respectively. The shaded component represents the range of validity of the parameters presented in Table \ref{tab:sedparams}. This is the range of values that encloses the 68 per cent probability (i.e. the shaded area represents 0.68 of the area under the curve) as defined in Equation \ref{eq:prob}. \label{fig:fit_sed}}
\end{figure*}

Figure~\ref{fig:fit_sed} presents the best fit SED (\emph{i.e.} the model with the smallest $\chi^2$ value) and the relative figures of merit (probability distribution diagrams) estimated from the Bayesian inference method for the temperature and mass of the cold component of the fit for three individual sources. These results were obtained from marginalization (\emph{i.e.} summing) of the probabilities of all models, where one parameter is fixed successively to its different values. The resulting histograms indicate the probability that a parameter takes a certain value, given the data and assumptions of our modelling. The width of the probability curves is a strong indicator of how well the data are constrained (see Section \ref{sec:method}).  The radius and temperature of the hot component are considered as ``nuisance parameters'' here. That is, they are parameters that have an influence on the data but do not have a direct physical interpretation and are thus not of prime interest to us in this study. 

 For each parameter $\theta$ with a density of probability $p(\theta)$, this range of validity is defined as the interval [$\theta_1$, $\theta_2$] where:
\begin{equation}\label{eq:prob}
 p(\theta_1) = p(\theta_2) \hspace{5mm} \mathrm{and} \hspace{5mm}\int_{\theta_1}^{\theta_2} p(\theta)\,\dd \theta = \gamma  
\end{equation}
with $\gamma = 0.68$. The interval [$\theta_1$, $\theta_2$] is a 68 per cent confidence interval and corresponds to the $1\,\sigma$ interval for a Gaussian distribution of probability. Table~\ref{tab:sedparams} gives the range of validity of the parameters for the cold component of the SED fit, with $min$ and $max$ representing the lower and upper values of this range, respectively.

\begin{table*}
\footnotesize
\caption[]{Parameters resulting from spectral energy distribution modelling of 180 of the 405 sources in the SIMBA source list. For each of the temperature, mass and luminosity, the range corresponding to a 1-sigma (68 per cent) probability of occurrence is presented, with the $_{min}$ and $_{max}$ values representing the lower and upper values of this range, respectively.\label{tab:sedparams} }
\centering
\begin{tabular}{@{}lllccrrrrrr@{}}
\hline 
\multicolumn{2}{c}{Peak position} & & Ident & Fit & \multicolumn{2}{c}{Temperature} & \multicolumn{2}{c}{Mass}& \multicolumn{2}{c}{Luminosity} \\
RA & Dec & Source Name & tracer & Type  & T$_{\rm cold}$$_{_{min}}$ &  T$_{\rm cold}$$_{_{max}}$ &  M$_{min}$ & M$_{max}$ & L$_{min}$ & L$_{max}$\\
(J2000) & (J2000) &  $^{a,b}$  &    $^c$    & $^d$ &  K   & K   &  \mstar     & \mstar & \lum & \lum\\
\hline
06 07 46.29 & -06 23 05.0 & G213.61-12.6  &  mr   &  SINGLE   &  2.8E+01  &  3.1E+01  &  9.3E+02  &  1.4E+03  &  7.6E+04   &  3.3E+05 \\
 06 09 06.50 & +21 50 42.0 & G188.79+1.02  &  r   &  SED   &  1.2E+01  &  3.4E+01  &  2.0E+02  &  1.4E+03  &  2.5E+02   &  2.9E+04 \\
 09 03 13.50 & -48 55 30.0 & G269.45-1.47  &  mr   &  SED   &  2.7E+00  &  3.5E+01  &  3.1E+02  &  7.6E+03  &  6.1E+02   &  7.0E+04 \\
 09 16 41.40 & -47 55 18.0 & G270.25+0.84  &  m   &  SED   &  2.7E+00  &  2.9E+01  &  1.5E+02  &  3.8E+03  &  8.9E+01   &  1.1E+04 \\
 10 48 03.98 & -58 27 04.0 & G287.37+0.65  &  m   &  SED   &  2.7E+00  &  3.6E+01  &  6.6E+01  &  2.2E+03  &  1.3E+02   &  2.0E+04 \\
 10 57 33.00 & -62 59 10.0 & G290.40-2.91  &  m   &  SED   &  2.7E+00  &  2.9E+01  &  6.6E+01  &  1.9E+03  &  3.7E+01   &  3.9E+03 \\
 11 12 18.06 & -58 46 19.0 & G290.37+1.66  &  m   &  SED   &  2.7E+00  &  3.5E+01  &  5.0E+01  &  1.4E+03  &  8.5E+00   &  1.1E+04 \\
 11 35 31.04 & -63 14 52.3 & G294.52-1.6$^\dag$  &  m   &  SED   &  2.7E+00  &  3.2E+01  &  1.1E+01  &  1.5E+02  &  1.1E+01   &  1.1E+03 \\
 12 11 47.64 & -61 46 29.5 & G298.26+0.7  &  m   &  SED   &  2.7E+00  &  3.6E+01  &  8.7E+01  &  2.5E+03  &  7.5E+01   &  1.7E+04 \\
 12 35 34.95 & -63 01 35.5 & G301.14-0.2  &  mr   &  SED   &  1.6E+01  &  3.9E+01  &  1.1E+03  &  3.1E+04  &  6.0E+02   &  2.5E+05 \\
 12 43 32.10 & -62 55 05.8 & G302.03-0.06  &  mr   &  SED   &  2.7E+00  &  3.6E+01  &  2.0E+02  &  5.0E+03  &  2.4E+02   &  4.5E+04 \\
 13 10 43.25 & -62 43 04.5 & G305.137+0.069$^\dag$  &  mm   &  SINGLE   &  3.4E+01  &  3.9E+01  &  8.7E+01  &  1.3E+02  &  2.7E+04   &  2.6E+05 \\
 13 16 58.38 & -62 55 25.2 & G305.833-0.196$^\dag$  &  mm   &  SINGLE   &  2.6E+01  &  2.8E+01  &  7.6E+01  &  1.0E+02  &  4.2E+03   &  1.8E+05 \\
 15 31 44.50 & -56 30 51.0 & G323.74-0.30$^\dag$  &  m   &  SED   &  2.7E+00  &  3.5E+01  &  2.7E+02  &  6.6E+03  &  3.7E+02   &  4.8E+04 \\
 16 11 26.90 & -51 41 57.0 & G331.279-0.189  &  m   &  SED   &  1.7E+01  &  3.6E+01  &  6.1E+02  &  1.3E+04  &  8.2E+02   &  1.4E+05 \\
 17 45 54.30 & -28 44 00.0 & G0.204+0.051$^\dag$  &  mm   &  SINGLE$^\alpha$   &  1.3E+01  &  3.5E+01  &  4.0E+02  &  1.9E+03  &  5.2E+02   &  2.0E+05 \\
 17 46 04.61 & -28 24 51.0 & G0.49+0.19$^\dag$  &  m   &  SED$^\alpha$   &  8.1E+00  &  3.2E+01  &  5.7E+01  &  6.1E+02  &  1.3E+01   &  2.7E+03 \\
 17 46 07.09 & -28 41 28.0 & G0.266-0.034$^\dag$  &  mm   &  SINGLE$^\alpha$   &  1.2E+01  &  3.2E+01  &  7.1E+02  &  3.3E+03  &  5.4E+02   &  2.1E+05 \\
 17 46 07.70 & -28 45 28.0 & G0.21-0.00$^\dag$  &  mr   &  SED$^\alpha$   &  8.4E+00  &  3.2E+01  &  7.1E+02  &  7.6E+03  &  9.8E+01   &  3.7E+04 \\
 17 46 08.24 & -28 25 23.0 & G0.497+0.170$^\dag$  &  mm   &  SED$^\alpha$   &  7.0E+00  &  2.9E+01  &  4.3E+01  &  5.3E+02  &  9.6E+00   &  1.8E+03 \\
 17 46 09.52 & -28 43 36.0 & G0.240+0.008$^\dag$  &  mm   &  SINGLE$^\alpha$   &  9.4E+00  &  2.4E+01  &  6.6E+03  &  3.1E+04  &  1.8E+03   &  2.9E+05 \\
 17 46 10.67 & -28 23 31.0 & G0.527+0.181$^\dag$  &  r   &  SED$^\alpha$   &  8.1E+00  &  3.2E+01  &  1.1E+02  &  1.2E+03  &  3.4E+01   &  6.6E+03 \\
 17 46 10.74 & -28 41 36.0 & G0.271+0.022$^\dag$  &  mm   &  SINGLE$^\alpha$   &  1.0E+01  &  3.4E+01  &  3.1E+02  &  1.9E+03  &  2.5E+02   &  1.8E+05 \\
 17 46 11.35 & -28 42 40.0 & G0.257+0.011$^\dag$  &  mm   &  SINGLE$^\alpha$   &  9.8E+00  &  2.5E+01  &  5.7E+03  &  2.7E+04  &  1.6E+03   &  2.9E+05 \\
 17 46 53.41 & -28 07 27.0 & G0.83+0.18$^\dag$  &  m   &  SED   &  1.3E+01  &  4.5E+01  &  1.0E+02  &  9.3E+02  &  1.0E+02   &  5.3E+04 \\
 17 47 01.19 & -28 45 36.0 & G0.310-0.170$^\dag$  &  mm   &  SINGLE$^\alpha$   &  1.6E+01  &  6.7E+01  &  4.3E+01  &  2.7E+02  &  3.8E+02   &  8.1E+05 \\
 17 47 09.71 & -28 46 08.0 & G0.32-0.20$^\dag$  &  mr   &  SINGLE$^\alpha$   &  9.8E+00  &  2.9E+01  &  3.3E+03  &  2.0E+04  &  1.4E+03   &  4.4E+05 \\
 17 47 20.66 & -28 46 56.0 & G0.325-0.242$^\dag$  &  mm   &  SINGLE$^\alpha$   &  1.3E+01  &  5.0E+01  &  1.3E+02  &  7.1E+02  &  3.6E+02   &  4.4E+05 \\
 17 48 31.59 & -28 00 30.9 & G1.124-0.065  &  mm   &  SINGLE   &  3.4E+01  &  5.6E+01  &  1.7E+02  &  4.0E+02  &  1.2E+04   &  7.4E+05 \\
 17 48 34.65 & -28 00 16.0 & G1.134-0.073  &  mm   &  SINGLE$^\alpha$   &  1.2E+01  &  5.8E+01  &  6.6E+01  &  5.3E+02  &  2.3E+02   &  7.2E+05 \\
 17 48 36.41 & -28 02 31.0 & G1.105-0.098  &  mm   &  SINGLE$^\alpha$   &  8.4E+00  &  3.0E+01  &  9.3E+02  &  7.6E+03  &  6.3E+02   &  4.0E+05 \\
 17 48 42.46 & -28 01 35.0 & G1.13-0.11  &  r   &  SED$^\alpha$   &  3.9E+00  &  2.0E+01  &  2.5E+03  &  9.3E+04  &  2.5E+02   &  2.0E+05 \\
 17 48 49.75 & -28 01 04.0 & G1.14-0.12  &  m   &  SINGLE$^\alpha$   &  1.5E+01  &  6.5E+01  &  7.6E+01  &  5.3E+02  &  3.4E+02   &  9.3E+05 \\
 17 50 15.11 & -27 54 23.0 & G0.55-0.85  &  mr   &  SINGLE$^\alpha$   &  9.4E+00  &  2.9E+01  &  6.1E+02  &  3.8E+03  &  2.5E+02   &  1.4E+05 \\
 17 50 18.77 & -28 53 19.0 & G0.549-0.868  &  mm   &  SINGLE$^\alpha$   &  1.4E+01  &  6.2E+01  &  7.1E+00  &  5.0E+01  &  4.9E+01   &  2.0E+05 \\
 17 50 25.46 & -28 50 15.0 & G0.627-0.848  &  mm   &  SINGLE$^\alpha$   &  1.5E+01  &  7.0E+01  &  5.3E+00  &  3.8E+01  &  3.4E+01   &  2.0E+05 \\
 17 50 26.07 & -28 52 31.0 & G0.600-0.871  &  mm   &  SINGLE$^\alpha$   &  1.0E+01  &  4.2E+01  &  1.9E+01  &  1.3E+02  &  4.0E+01   &  1.4E+05 \\
 17 50 46.50 & -26 39 44.0 & G2.54+0.20$^\dag$  &  m   &  SINGLE   &  1.9E+01  &  2.7E+01  &  1.7E+02  &  4.0E+02  &  5.6E+02   &  1.4E+05 \\
 17 59 02.84 & -24 20 55.0 & G5.48-0.24  &  r   &  SED$^\alpha$   &  8.4E+00  &  3.4E+01  &  1.4E+03  &  1.5E+04  &  5.3E+02   &  1.9E+05 \\
 17 59 07.53 & -24 19 19.0 & G5.504-0.246  &  mm   &  SINGLE$^\alpha$   &  9.8E+00  &  3.2E+01  &  8.1E+02  &  5.0E+03  &  8.5E+02   &  4.1E+05 \\
 18 00 30.42 & -24 03 59.0 & G5.89-0.39  &  r   &  SED$^\alpha$   &  4.1E+00  &  1.9E+01  &  7.1E+02  &  2.3E+04  &  9.6E+01   &  4.1E+04 \\
 18 00 40.90 & -24 04 12.0 & G5.90-0.42$^\dag$  &  m   &  SED$^\alpha$   &  9.4E+00  &  4.0E+01  &  3.5E+02  &  3.8E+03  &  1.3E+02   &  6.6E+04 \\
 18 00 49.74 & -23 20 25.0 & G6.53-0.10  &  r   &  SED   &  1.3E+01  &  3.5E+01  &  3.3E+03  &  2.7E+04  &  5.6E+03   &  6.6E+05 \\
 18 00 54.58 & -23 16 54.0 & G6.60-0.08$^\dag$  &  m   &  SED$^\alpha$   &  4.9E+00  &  1.2E+01  &  1.0E+00  &  4.0E+00  &  4.9E-02   &  4.2E+01 \\
 18 02 49.31 & -21 48 34.0 & G8.111+0.257  &  mm   &  SINGLE$^\alpha$   &  1.0E+01  &  6.2E+01  &  6.1E+00  &  7.6E+01  &  1.9E+01   &  2.1E+05 \\
 18 02 52.76 & -21 47 54.0 & G8.127+0.255  &  mm   &  SINGLE$^\alpha$   &  5.9E+00  &  3.1E+01  &  3.8E+01  &  6.1E+02  &  2.2E+01   &  1.3E+05 \\
 18 02 56.21 & -21 47 38.0 & G8.138+0.246  &  mm   &  SINGLE$^\alpha$   &  6.8E+00  &  3.2E+01  &  1.0E+02  &  1.4E+03  &  4.4E+01   &  1.3E+05 \\
 18 03 01.95 & -21 48 02.0 & G8.13+0.22  &  mr   &  SED$^\alpha$   &  6.5E+00  &  3.0E+01  &  5.3E+02  &  1.0E+04  &  8.1E+01   &  4.0E+04 \\
 18 03 26.85 & -24 22 29.0 & G5.948-1.125  &  mm   &  SINGLE$^\alpha$   &  7.8E+00  &  4.7E+01  &  3.5E+00  &  5.0E+01  &  7.8E+00   &  1.1E+05 \\
 18 03 29.19 & -24 21 49.0 & G5.962-1.128  &  mm   &  SINGLE$^\alpha$   &  7.3E+00  &  3.2E+01  &  7.1E+00  &  8.7E+01  &  9.3E+00   &  8.3E+04 \\
 18 03 33.88 & -24 21 41.0 & G5.975-1.146  &  mm   &  SINGLE$^\alpha$   &  8.7E+00  &  5.4E+01  &  4.6E+00  &  5.7E+01  &  1.2E+01   &  1.5E+05 \\
 18 03 36.80 & -24 22 08.0 & G5.971-1.158  &  mm   &  SINGLE$^\alpha$   &  1.3E+01  &  5.4E+01  &  1.4E+01  &  1.0E+02  &  7.0E+01   &  2.1E+05 \\
 18 03 41.49 & -24 22 37.0 & G5.97-1.17  &  r   &  SED$^\alpha$   &  3.2E+00  &  1.7E+01  &  1.3E+02  &  5.7E+03  &  1.6E+01   &  2.0E+04 \\
 18 05 13.33 & -18 50 30.0 & G10.10+0.72$^\dag$  &  r   &  SED   &  5.3E+00  &  2.0E+01  &  1.0E+00  &  7.1E+00  &  1.3E-01   &  3.3E+01 \\
 18 06 14.80 & -20 31 29.0 & G9.63+0.19  &  mr   &  SED$^\alpha$   &  7.8E+00  &  3.1E+01  &  2.7E+02  &  2.8E+03  &  5.2E+01   &  1.3E+04 \\
 18 06 18.91 & -21 37 21.0 & G8.68-0.36  &  mr   &  SINGLE$^\alpha$   &  8.4E+00  &  2.5E+01  &  3.3E+03  &  2.0E+04  &  7.3E+02   &  2.2E+05 \\
 18 06 23.49 & -21 36 57.0 & G8.686-0.366  &  m   &  SINGLE$^\alpha$   &  1.1E+01  &  3.1E+01  &  7.1E+02  &  3.8E+03  &  4.7E+02   &  2.0E+05 \\
 18 07 50.36 & -20 18 51.0 & G9.99-0.03  &  m   &  SINGLE$^\alpha$   &  8.7E+00  &  2.7E+01  &  3.1E+02  &  1.9E+03  &  9.8E+01   &  9.7E+04 \\
 18 07 53.21 & -20 18 19.0 & G10.001-0.033  &  r   &  SED$^\alpha$   &  7.0E+00  &  3.4E+01  &  5.0E+01  &  8.1E+02  &  9.9E+00   &  2.5E+03 \\
 18 08 38.47 & -19 51 48.0 & G10.47+0.02$^\dag$  &  mr   &  SED$^\alpha$   &  7.3E+00  &  2.6E+01  &  7.6E+03  &  9.3E+04  &  7.1E+02   &  1.5E+05 \\
 18 08 45.47 & -19 54 30.0 & G10.44-0.01$^\dag$  &  m   &  SINGLE$^\alpha$   &  1.3E+01  &  4.2E+01  &  3.5E+02  &  1.9E+03  &  6.4E+02   &  4.3E+05 \\
\end{tabular}
\end{table*}

\begin{table*}
\footnotesize
\contcaption
\centering
\begin{tabular}{@{}lllccrrrrrr@{}}
\hline 
\multicolumn{2}{c}{Peak position} & & Ident & Fit & \multicolumn{2}{c}{Temperature} & \multicolumn{2}{c}{Mass}& \multicolumn{2}{c}{Luminosity} \\
RA & Dec & Source Name & tracer & Type  & T$_{\rm cold}$$_{_{min}}$ &  T$_{\rm cold}$$_{_{max}}$ &  M$_{min}$ & M$_{max}$ & L$_{min}$ & L$_{max}$\\
(J2000) & (J2000) &  $^{a,b}$  &    $^c$    & $^d$ &  K   & K   &  \mstar     & \mstar & \lum & \lum\\
\hline 
18 08 45.85 & -20 05 42.0 & G10.287-0.110$^\dag$  &  mm   &  SINGLE$^\alpha$   &  7.8E+00  &  2.6E+01  &  8.7E+01  &  6.1E+02  &  6.1E+01   &  9.4E+04 \\
 18 08 49.25 & -20 05 58.0 & G10.284-0.126  &  m   &  SED$^\alpha$   &  7.8E+00  &  3.4E+01  &  5.7E+01  &  7.1E+02  &  2.0E+01   &  7.1E+03 \\
 18 08 52.66 & -20 05 58.0 & G10.288-0.127$^\dag$  &  mm   &  SINGLE$^\alpha$   &  1.3E+01  &  4.8E+01  &  1.6E+01  &  1.0E+02  &  6.2E+01   &  1.4E+05 \\
 18 08 56.07 & -20 05 50.0 & G10.29-0.14  &  mr   &  SED$^\alpha$   &  7.3E+00  &  2.9E+01  &  2.0E+02  &  2.5E+03  &  5.2E+01   &  1.5E+04 \\
 18 09 00.04 & -20 03 34.0 & G10.343-0.142$^\dag$  &  m   &  SINGLE$^\alpha$   &  1.3E+01  &  4.2E+01  &  4.3E+01  &  2.0E+02  &  8.2E+01   &  1.1E+05 \\
 18 09 03.49 & -20 02 54.0 & G10.359-0.149$^\dag$  &  mm   &  SINGLE$^\alpha$   &  1.1E+01  &  4.7E+01  &  2.2E+01  &  1.7E+02  &  3.7E+01   &  1.2E+05 \\
 18 09 21.03 & -20 19 25.0 & G10.15-0.34  &  r   &  SINGLE$^\alpha$   &  3.8E+00  &  1.9E+01  &  9.3E+02  &  2.3E+04  &  1.8E+02   &  4.1E+05 \\
 18 10 15.59 & -19 54 45.0 & G10.63-0.33B  &  mm   &  SINGLE$^\alpha$   &  1.1E+01  &  5.2E+01  &  2.0E+02  &  1.9E+03  &  5.5E+02   &  1.1E+06 \\
 18 10 18.42 & -19 54 29.0 & G10.62-0.33  &  m   &  SED$^\alpha$   &  7.0E+00  &  3.0E+01  &  8.1E+02  &  1.1E+04  &  2.0E+02   &  4.6E+04 \\
 18 10 19.00 & -20 45 25.0 & G9.88-0.75  &  r   &  SED$^\alpha$   &  7.8E+00  &  2.6E+01  &  8.1E+02  &  7.6E+03  &  8.9E+01   &  1.6E+04 \\
 18 10 23.56 & -20 43 09.0 & G9.924-0.749  &  mm   &  SED$^\alpha$   &  4.9E+00  &  2.5E+01  &  5.7E+01  &  1.1E+03  &  1.6E+01   &  7.3E+03 \\
 18 10 28.77 & -19 55 48.0 & G10.62-0.38  &  mr   &  SINGLE$^\alpha$   &  8.7E+00  &  2.8E+01  &  8.7E+03  &  6.1E+04  &  2.3E+03   &  8.0E+05 \\
 18 11 23.87 & -19 32 20.0 & G11.075-0.384  &  mm   &  SED$^\alpha$   &  4.9E+00  &  2.5E+01  &  5.7E+01  &  1.1E+03  &  1.6E+01   &  7.3E+03 \\
 18 11 31.80 & -19 30 44.0 & G11.11-0.34  &  r   &  SED$^\alpha$   &  8.1E+00  &  2.8E+01  &  8.1E+02  &  7.6E+03  &  1.2E+02   &  2.8E+04 \\
 18 11 35.76 & -19 30 44.0 & G11.117-0.413  &  mm   &  SINGLE$^\alpha$   &  1.2E+01  &  3.6E+01  &  1.3E+02  &  6.1E+02  &  1.3E+02   &  1.2E+05 \\
 18 11 51.40 & -17 31 30.0 & G12.88+0.48$^\dag$  &  m   &  SINGLE$^\alpha$   &  8.1E+00  &  2.6E+01  &  1.1E+03  &  7.6E+03  &  2.6E+02   &  1.4E+05 \\
 18 11 52.90 & -18 36 03.0 & G11.948-0.003$^\dag$  &  mm   &  SED   &  2.7E+00  &  4.3E+01  &  1.5E+02  &  3.8E+03  &  6.9E+02   &  1.2E+05 \\
 18 11 53.64 & -17 30 02.0 & G12.914+0.493$^\dag$  &  mm   &  SINGLE$^\alpha$   &  6.8E+00  &  2.9E+01  &  6.6E+01  &  7.1E+02  &  3.5E+01   &  9.0E+04 \\
 18 12 01.34 & -18 31 55.0 & G12.02-0.03$^\dag$  &  m   &  SED   &  1.4E+01  &  2.7E+01  &  2.3E+02  &  9.3E+02  &  2.2E+02   &  1.1E+04 \\
 18 12 11.11 & -18 41 30.0 & G11.903-0.140$^\dag$  &  mr   &  SINGLE$^\alpha$   &  4.7E+00  &  1.7E+01  &  3.5E+02  &  5.7E+03  &  3.3E+01   &  6.5E+04 \\
 18 12 15.61 & -18 44 58.0 & G11.861-0.183$^\dag$  &  mm   &  SINGLE$^\alpha$   &  1.4E+01  &  5.6E+01  &  1.2E+01  &  7.6E+01  &  5.3E+01   &  1.5E+05 \\
 18 12 17.30 & -18 40 02.0 & G11.93-0.14$^\dag$  &  m   &  SINGLE$^\alpha$   &  1.0E+01  &  2.6E+01  &  1.5E+02  &  7.1E+02  &  8.5E+01   &  8.5E+04 \\
 18 12 19.55 & -18 39 54.0 & G11.942-0.157$^\dag$  &  mm   &  SINGLE$^\alpha$   &  8.7E+00  &  3.4E+01  &  7.6E+01  &  7.1E+02  &  5.0E+01   &  9.6E+04 \\
 18 12 33.13 & -18 30 05.0 & G12.112-0.125  &  mm   &  SINGLE   &  2.5E+01  &  2.7E+01  &  1.2E+03  &  1.6E+03  &  4.1E+04   &  2.8E+05 \\
 18 12 39.31 & -18 24 13.0 & G12.20-0.09  &  mr   &  SINGLE$^\alpha$   &  1.0E+01  &  3.0E+01  &  8.7E+03  &  4.6E+04  &  4.2E+03   &  9.2E+05 \\
 18 12 43.25 & -18 25 09.0 & G12.18-0.12A  &  m   &  SINGLE$^\alpha$   &  1.3E+01  &  4.3E+01  &  8.1E+02  &  5.0E+03  &  1.3E+03   &  8.9E+05 \\
 18 12 44.37 & -18 24 21.0 & G12.216-0.119  &  mm   &  SINGLE$^\alpha$   &  1.2E+01  &  3.6E+01  &  1.9E+03  &  1.0E+04  &  2.6E+03   &  8.4E+05 \\
 18 12 50.64 & -18 40 31.0 & G11.99-0.27$^\dag$  &  m   &  SED$^\alpha$   &  8.7E+00  &  3.2E+01  &  7.6E+01  &  7.1E+02  &  1.5E+01   &  3.6E+03 \\
 18 12 54.72 & -18 11 04.0 & G12.43-0.05  &  r   &  SED   &  1.1E+01  &  3.0E+01  &  1.9E+03  &  1.3E+04  &  1.1E+03   &  1.6E+05 \\
 18 13 54.14 & -18 01 41.0 & G12.68-0.18$^\dag$  &  m   &  SED$^\alpha$   &  7.3E+00  &  2.8E+01  &  9.3E+02  &  8.7E+03  &  2.5E+02   &  2.8E+04 \\
 18 13 58.08 & -18 54 14.0 & G11.94-0.62B  &  mm   &  SINGLE$^\alpha$   &  6.8E+00  &  1.9E+01  &  7.1E+02  &  5.0E+03  &  1.6E+02   &  1.0E+05 \\
 18 14 00.90 & -18 53 18.0 & G11.93-0.61  &  mr   &  SED$^\alpha$   &  7.8E+00  &  3.0E+01  &  6.1E+02  &  7.6E+03  &  7.4E+01   &  2.8E+04 \\
 18 14 07.04 & -18 00 37.0 & G12.722-0.218$^\dag$  &  mm   &  SED$^\alpha$   &  5.9E+00  &  2.1E+01  &  3.1E+02  &  4.3E+03  &  9.7E+01   &  1.6E+04 \\
 18 14 33.90 & -17 51 44.0 & G12.90-0.25B$^\dag$  &  mm   &  SINGLE$^\alpha$   &  8.4E+00  &  3.6E+01  &  1.3E+02  &  1.2E+03  &  6.1E+01   &  1.3E+05 \\
 18 14 35.54 & -16 45 36.0 & G13.87+0.28  &  m   &  SED   &  9.1E+00  &  3.7E+01  &  5.3E+02  &  7.6E+03  &  3.0E+02   &  1.2E+05 \\
 18 14 36.13 & -17 54 56.0 & G12.859-0.272$^\dag$  &  mm   &  SED$^\alpha$   &  6.5E+00  &  3.2E+01  &  1.7E+02  &  3.3E+03  &  3.4E+01   &  1.0E+04 \\
 18 14 38.94 & -17 51 52.0 & G12.90-0.26$^\dag$  &  m   &  SED$^\alpha$   &  7.3E+00  &  2.9E+01  &  9.3E+02  &  1.3E+04  &  1.3E+02   &  3.3E+04 \\
 18 16 22.10 & -19 41 19.0 & G11.49-1.48$^\dag$  &  m   &  SED   &  1.1E+01  &  2.8E+01  &  4.3E+01  &  3.1E+02  &  1.3E+01   &  2.2E+03 \\
 18 17 02.17 & -16 14 28.0 & G14.60+0.01$^\dag$  &  mr   &  SED$^\alpha$   &  8.1E+00  &  2.9E+01  &  1.5E+02  &  1.6E+03  &  2.1E+01   &  4.3E+03 \\
 18 19 12.03 & -20 47 23.0 & G10.84-2.59  &  r   &  SINGLE$^\alpha$   &  7.8E+00  &  2.5E+01  &  1.5E+02  &  1.1E+03  &  6.6E+01   &  8.3E+04 \\
 18 20 23.10 & -16 11 31.0 & G15.03-0.67$^\dag$  &  mr   &  SINGLE$^\alpha$   &  8.4E+00  &  3.4E+01  &  1.1E+03  &  1.0E+04  &  4.7E+02   &  4.0E+05 \\
 18 21 09.10 & -14 31 40.0 & G16.58-0.05$^\dag$  &  m   &  SED$^\alpha$   &  7.6E+00  &  2.6E+01  &  6.1E+02  &  6.6E+03  &  3.4E+01   &  9.4E+03 \\
 18 21 14.61 & -14 32 52.0 & G16.580-0.079$^\dag$  &  mm   &  SINGLE$^\alpha$   &  5.1E+00  &  1.0E+01  &  4.0E+02  &  2.2E+03  &  1.2E+01   &  3.2E+04 \\
 18 25 01.30 & -13 15 27.0 & G18.15-0.28  &  r   &  SINGLE$^\alpha$   &  3.2E+00  &  1.7E+01  &  5.7E+01  &  2.2E+03  &  1.1E+01   &  8.7E+04 \\
 18 25 07.33 & -13 14 23.0 & G18.177-0.296  &  mm   &  SINGLE$^\alpha$   &  1.0E+01  &  4.8E+01  &  2.8E+01  &  2.7E+02  &  4.7E+01   &  1.6E+05 \\
 18 25 41.65 & -13 10 16.0 & G18.30-0.39  &  r   &  SED   &  9.8E+00  &  3.5E+01  &  2.0E+02  &  3.3E+03  &  6.5E+01   &  3.3E+04 \\
 18 27 16.34 & -11 53 51.0 & G19.61-0.1$^\dag$  &  m   &  SED$^\alpha$   &  8.4E+00  &  3.4E+01  &  1.5E+02  &  1.9E+03  &  1.4E+01   &  7.8E+03 \\
 18 27 37.86 & -11 56 40.0 & G19.607-0.234$^\dag$  &  mr   &  SED   &  7.8E+00  &  3.5E+01  &  5.3E+02  &  1.3E+04  &  1.7E+02   &  7.9E+04 \\
 18 27 55.30 & -11 52 48.0 & G19.70-0.27A$^\dag$  &  m   &  SED   &  1.7E+01  &  3.9E+01  &  5.3E+02  &  1.3E+04  &  1.6E+03   &  2.3E+05 \\
 18 29 24.20 & -15 16 06.0 & G16.86-2.15$^\dag$  &  m   &  SED$^\alpha$   &  6.8E+00  &  1.8E+01  &  7.1E+02  &  5.0E+03  &  5.6E+01   &  2.6E+03 \\
 18 29 33.60 & -15 15 50.0 & G16.883-2.188$^\dag$  &  mm   &  SINGLE$^\alpha$   &  1.0E+01  &  3.2E+01  &  1.1E+01  &  6.6E+01  &  9.4E+00   &  4.0E+04 \\
 18 31 02.64 & -09 49 38.0 & G21.87+0.01$^\dag$  &  mr   &  SED$^\alpha$   &  8.4E+00  &  3.2E+01  &  2.8E+01  &  2.7E+02  &  8.8E+00   &  2.1E+03 \\
 18 31 43.02 & -09 22 28.0 & G22.36+0.07  &  m   &  SED$^\alpha$   &  2.7E+00  &  1.3E+01  &  1.7E+02  &  1.3E+04  &  1.4E+01   &  3.3E+03 \\
 18 33 53.06 & -08 07 23.0 & G23.71+0.17  &  r   &  SED$^\alpha$   &  7.3E+00  &  2.8E+01  &  1.1E+03  &  1.1E+04  &  1.7E+02   &  5.9E+04 \\
 18 33 53.60 & -08 08 51.0 & G23.689+0.159  &  mm   &  SINGLE$^\alpha$   &  1.2E+01  &  4.7E+01  &  8.7E+01  &  4.6E+02  &  2.1E+02   &  2.9E+05 \\
 18 34 09.23 & -07 17 45.0 & G24.47+0.49$^\dag$  &  r   &  SED$^\alpha$   &  9.1E+00  &  3.7E+01  &  8.1E+02  &  1.0E+04  &  1.9E+02   &  1.9E+05 \\
 18 34 20.90 & -05 59 48.0 & G25.65+1.04  &  mr   &  SED$^\alpha$   &  7.8E+00  &  2.6E+01  &  6.1E+02  &  5.7E+03  &  8.1E+01   &  1.3E+04 \\
 18 34 31.30 & -08 42 47.0 & G23.25-0.24$^\dag$  &  m   &  SINGLE$^\alpha$   &  1.0E+01  &  2.8E+01  &  7.6E+01  &  3.5E+02  &  3.5E+01   &  6.6E+04 \\
 18 34 36.16 & -08 42 39.0 & G23.268-0.257$^\dag$  &  mm   &  SINGLE$^\alpha$   &  5.4E+00  &  1.7E+01  &  8.1E+02  &  8.7E+03  &  5.4E+01   &  7.2E+04 \\
 18 34 39.20 & -08 31 41.0 & G23.43-0.18$^\dag$  &  m   &  SINGLE$^\alpha$   &  1.0E+01  &  2.5E+01  &  1.9E+03  &  8.7E+03  &  7.6E+02   &  1.8E+05 \\
 18 36 06.69 & -07 13 47.0 & G23.754+0.095$^\dag$  &  mm   &  SINGLE$^\alpha$   &  9.1E+00  &  3.2E+01  &  3.1E+02  &  2.5E+03  &  1.6E+02   &  1.5E+05 \\
 18 36 12.60 & -07 12 11.0 & G24.78+0.08$^\dag$  &  m   &  SINGLE$^\alpha$   &  7.8E+00  &  2.3E+01  &  5.7E+03  &  4.0E+04  &  9.5E+02   &  2.4E+05 \\
 18 36 17.86 & -07 08 52.0 & G24.84+0.08$^\dag$  &  m   &  SINGLE$^\alpha$   &  1.0E+01  &  3.4E+01  &  5.3E+02  &  3.3E+03  &  3.6E+02   &  2.0E+05 \\
\end{tabular}
\end{table*}

\begin{table*}
\footnotesize
\contcaption
\centering
\begin{tabular}{@{}lllccrrrrrr@{}}
\hline 
\multicolumn{2}{c}{Peak position} & & Ident & Fit & \multicolumn{2}{c}{Temperature} & \multicolumn{2}{c}{Mass}& \multicolumn{2}{c}{Luminosity} \\
RA & Dec & Source Name & tracer & Type  & T$_{\rm cold}$$_{_{min}}$ &  T$_{\rm cold}$$_{_{max}}$ &  M$_{min}$ & M$_{max}$ & L$_{min}$ & L$_{max}$\\
(J2000) & (J2000) &  $^{a,b}$  &    $^c$    & $^d$ &  K   & K   &  \mstar     & \mstar & \lum & \lum\\
\hline
18 36 25.92 & -07 05 16.9 & G24.919+0.088$^\dag$  &  mm   &  SED   &  9.8E+00  &  3.2E+01  &  8.1E+02  &  8.7E+03  &  2.9E+02   &  8.5E+04 \\
 18 38 03.00 & -06 24 09.0 & G25.70+0.04  &  mr   &  SED$^\alpha$   &  7.0E+00  &  3.0E+01  &  1.9E+03  &  2.7E+04  &  2.3E+02   &  7.9E+04 \\
 18 39 03.94 & -06 24 13.0 & G25.82-0.17$^\dag$  &  m   &  SINGLE$^\alpha$   &  8.7E+00  &  2.3E+01  &  1.9E+03  &  1.0E+04  &  4.4E+02   &  1.5E+05 \\
 18 42 42.60 & -04 15 39.0 & G28.14-0.00$^\dag$  &  m   &  SED   &  1.3E+01  &  3.7E+01  &  2.0E+02  &  1.4E+03  &  1.0E+02   &  3.9E+04 \\
 18 42 54.89 & -04 07 40.0 & G28.287+0.010$^\dag$  &  mm   &  SINGLE$^\alpha$   &  1.0E+01  &  3.9E+01  &  1.0E+02  &  8.1E+02  &  1.4E+02   &  2.1E+05 \\
 18 42 58.10 & -04 13 56.0 & G28.20-0.04$^\dag$ &  mr   &  SED$^\alpha$   &  7.8E+00  &  2.8E+01  &  2.2E+03  &  2.7E+04  &  2.5E+02   &  6.3E+04 \\
 18 43 02.91 & -04 14 52.0 & G29.193-0.073$^\dag$  &  mm   &  SINGLE$^\alpha$   &  1.0E+01  &  4.3E+01  &  6.6E+01  &  5.3E+02  &  1.3E+02   &  2.4E+05 \\
 18 44 15.17 & -04 01 56.0 & G28.28-0.35  &  mr   &  SED$^\alpha$   &  5.7E+00  &  2.9E+01  &  1.7E+02  &  4.3E+03  &  3.2E+01   &  2.2E+04 \\
 18 44 21.57 & -04 17 35.0 & G28.31-0.38$^\dag$  &  m   &  SED$^\alpha$   &  9.4E+00  &  4.3E+01  &  1.3E+02  &  1.6E+03  &  8.2E+01   &  7.1E+04 \\
 18 45 52.76 & -02 42 29.0 & G29.888+0.001$^\dag$  &  mm   &  SINGLE$^\alpha$   &  5.4E+00  &  2.3E+01  &  2.7E+02  &  4.3E+03  &  1.1E+02   &  3.3E+05 \\
 18 45 54.36 & -02 42 45.0 & G29.889-0.006$^\dag$  &  mm   &  SINGLE$^\alpha$   &  5.1E+00  &  1.6E+01  &  2.3E+02  &  2.8E+03  &  3.0E+01   &  1.2E+05 \\
 18 45 59.70 & -02 41 17.0 & G29.918-0.014$^\dag$  &  mm   &  SINGLE$^\alpha$   &  6.8E+00  &  3.0E+01  &  6.6E+01  &  8.1E+02  &  7.7E+01   &  3.8E+05 \\
 18 46 00.23 & -02 45 09.0 & G29.86-0.04$^\dag$  &  m   &  SED$^\alpha$   &  1.0E+01  &  4.2E+01  &  2.7E+02  &  2.5E+03  &  1.1E+02   &  4.3E+04 \\
 18 46 01.30 & -02 45 25.0 & G29.861-0.053$^\dag$  &  mm   &  SINGLE$^\alpha$   &  1.0E+01  &  5.6E+01  &  7.6E+01  &  8.1E+02  &  2.8E+02   &  1.0E+06 \\
 18 46 02.37 & -02 45 57.0 & G29.853-0.062$^\dag$  &  mm   &  SINGLE$^\alpha$   &  1.2E+01  &  5.4E+01  &  1.3E+02  &  1.1E+03  &  5.0E+02   &  1.0E+06 \\
 18 46 03.97 & -02 39 25.0 & G29.96-0.02B$^\dag$  &  mr   &  SED$^\alpha$   &  7.3E+00  &  3.4E+01  &  1.9E+03  &  3.1E+04  &  2.2E+02   &  1.6E+05 \\
 18 46 05.04 & -02 42 29.0 & G29.912-0.045$^\dag$  &  mm   &  SINGLE$^\alpha$   &  1.1E+01  &  3.7E+01  &  8.1E+02  &  5.0E+03  &  1.3E+03   &  6.3E+05 \\
 18 46 06.11 & -02 41 25.0 & G29.930-0.040$^\dag$  &  mm   &  SINGLE$^\alpha$   &  1.2E+01  &  5.8E+01  &  6.6E+01  &  5.3E+02  &  3.5E+02   &  1.1E+06 \\
 18 46 09.84 & -02 41 25.0 & G29.937-0.054$^\dag$  &  mm   &  SED$^\alpha$   &  3.5E+00  &  1.8E+01  &  1.3E+02  &  5.7E+03  &  2.4E+01   &  1.6E+04 \\
 18 46 11.45 & -02 42 05.0 & G29.945-0.059  &  mm   &  SINGLE$^\alpha$   &  3.2E+00  &  2.0E+01  &  2.0E+02  &  8.7E+03  &  5.7E+01   &  4.1E+05 \\
 18 46 12.51 & -02 39 09.0 & G29.978-0.050$^\dag$  &  m   &  SINGLE$^\alpha$   &  1.2E+01  &  4.2E+01  &  4.6E+02  &  2.5E+03  &  1.1E+03   &  7.2E+05 \\
 18 46 58.62 & -02 07 27.0 & G30.533-0.023$^\dag$  &  mm   &  SED$^\alpha$   &  9.1E+00  &  4.3E+01  &  2.8E+01  &  3.5E+02  &  5.1E+00   &  7.3E+03 \\
 18 47 06.97 & -01 46 42.0 & G30.855+0.149$^\dag$  &  mm   &  SINGLE   &  2.7E+01  &  3.0E+01  &  6.1E+02  &  9.3E+02  &  3.9E+04   &  2.8E+05 \\
 18 47 08.57 & -01 44 02.0 & G30.89+0.16$^\dag$  &  m   &  SED   &  1.3E+01  &  3.7E+01  &  2.0E+02  &  1.4E+03  &  1.0E+02   &  3.9E+04 \\
 18 47 13.37 & -01 44 58.0 & G30.894+0.140$^\dag$  &  mm   &  SINGLE$^\alpha$   &  8.4E+00  &  2.8E+01  &  4.0E+02  &  2.8E+03  &  1.5E+02   &  1.2E+05 \\
 18 47 15.50 & -01 47 06.0 & G30.869+0.116$^\dag$  &  r   &  SED   &  2.7E+00  &  3.4E+01  &  5.3E+02  &  1.5E+04  &  5.1E+02   &  7.5E+04 \\
 18 47 18.37 & -02 06 15.0 & G30.59-0.04$^\dag$  &  m   &  SED   &  1.2E+01  &  3.6E+01  &  1.7E+02  &  1.2E+03  &  7.6E+01   &  2.4E+04 \\
 18 47 26.71 & -01 44 50.0 & G30.924+0.092$^\dag$  &  mm   &  SINGLE   &  1.6E+01  &  3.6E+01  &  1.1E+02  &  7.1E+02  &  9.8E+01   &  1.2E+05 \\
 18 47 34.77 & -01 12 47.0 & G31.41+0.30  &  mr   &  SED$^\alpha$   &  4.2E+00  &  1.3E+01  &  1.1E+04  &  2.5E+05  &  3.7E+02   &  1.9E+04 \\
 18 47 34.90 & -01 56 41.0 & G30.760-0.027$^\dag$  &  mm   &  SED$^\alpha$   &  3.8E+00  &  1.9E+01  &  3.1E+02  &  7.6E+03  &  6.2E+01   &  1.0E+05 \\
 18 47 35.43 & -02 01 59.0 & G30.682-0.072$^\dag$  &  mm   &  SINGLE$^\alpha$   &  7.3E+00  &  3.0E+01  &  5.3E+02  &  5.7E+03  &  1.4E+02   &  1.5E+05 \\
 18 47 35.80 & -01 55 29.0 & G30.78-0.02  &  m   &  SED$^\alpha$   &  6.1E+00  &  2.3E+01  &  1.6E+03  &  2.7E+04  &  1.4E+02   &  1.2E+05 \\
 18 47 35.97 & -02 01 03.0 & G30.705-0.065$^\dag$  &  m   &  SINGLE$^\alpha$   &  9.4E+00  &  2.6E+01  &  3.3E+03  &  1.7E+04  &  9.2E+02   &  2.3E+05 \\
 18 47 38.10 & -01 57 45.0 & G30.76-0.05$^\dag$  &  mm   &  SED   &  1.7E+01  &  5.6E+01  &  3.5E+02  &  1.9E+03  &  1.5E+04   &  1.1E+06 \\
 18 47 39.17 & -01 58 41.0 & G30.740-0.060$^\dag$  &  mm   &  SINGLE$^\alpha$   &  8.4E+00  &  2.9E+01  &  9.3E+02  &  7.6E+03  &  2.8E+02   &  1.7E+05 \\
 18 47 41.30 & -02 00 33.0 & G30.716-0.082$^\dag$  &  mm   &  SED$^\alpha$   &  5.4E+00  &  2.5E+01  &  7.1E+02  &  2.0E+04  &  5.5E+01   &  1.6E+04 \\
 18 47 41.83 & -01 59 45.0 & G30.729-0.078$^\dag$  &  mm   &  SINGLE$^\alpha$   &  1.0E+01  &  3.4E+01  &  3.5E+02  &  2.2E+03  &  2.6E+02   &  1.6E+05 \\
 18 47 45.94 & -01 54 25.0 & G30.81-0.05$^\dag$  &  m   &  SINGLE$^\alpha$   &  9.1E+00  &  2.3E+01  &  8.7E+03  &  4.6E+04  &  2.0E+03   &  3.2E+05 \\
 18 48 01.58 & -01 36 01.0 & G31.119+0.029$^\dag$  &  mm   &  SINGLE$^\alpha$   &  9.4E+00  &  3.4E+01  &  2.3E+00  &  1.6E+01  &  3.1E+00   &  2.4E+04 \\
 18 48 10.23 & -01 27 58.0 & G31.256+0.061  &  mm   &  SINGLE$^\alpha$   &  1.0E+01  &  3.2E+01  &  2.7E+02  &  1.6E+03  &  2.8E+02   &  1.7E+05 \\
 18 48 11.87 & -01 26 22.0 & G31.28+0.06  &  mr   &  SED$^\alpha$   &  7.3E+00  &  2.8E+01  &  1.4E+03  &  2.0E+04  &  1.6E+02   &  3.6E+04 \\
 18 49 32.57 & -01 28 56.0 & G31.40-0.26  &  r   &  SED$^\alpha$   &  7.8E+00  &  3.2E+01  &  1.1E+03  &  1.3E+04  &  2.1E+02   &  9.0E+04 \\
 18 49 34.17 & -01 29 44.0 & G31.388-0.266  &  mm   &  SINGLE$^\alpha$   &  1.4E+01  &  6.0E+01  &  3.3E+01  &  2.3E+02  &  2.1E+02   &  5.6E+05 \\
 18 50 30.70 & -00 02 00.0 & G32.80+0.19  &  r   &  SED$^\alpha$   &  7.0E+00  &  2.9E+01  &  1.0E+04  &  1.4E+05  &  1.4E+03   &  5.0E+05 \\
 18 52 08.00 & +00 08 10.0 & G33.13-0.09  &  mr   &  SINGLE$^\alpha$   &  1.1E+01  &  2.6E+01  &  1.2E+03  &  5.0E+03  &  6.4E+02   &  1.6E+05 \\
 18 52 50.73 & +00 55 28.0 & G33.92+0.11  &  r   &  SED$^\alpha$   &  7.3E+00  &  2.7E+01  &  3.3E+03  &  4.0E+04  &  4.4E+02   &  1.1E+05 \\
 18 53 17.97 & +01 14 57.0 & G34.256+0.150  &  m   &  SINGLE$^\alpha$   &  9.8E+00  &  2.7E+01  &  6.6E+03  &  3.5E+04  &  2.5E+03   &  5.0E+05 \\
 18 53 59.97 & +02 01 08.0 & G35.02+0.35  &  mr   &  SED$^\alpha$   &  5.4E+00  &  2.5E+01  &  7.1E+02  &  2.0E+04  &  5.5E+01   &  1.6E+04 \\
 18 56 00.67 & +02 22 51.0 & G35.57+0.07  &  r   &  SED$^\alpha$   &  7.6E+00  &  3.2E+01  &  9.3E+02  &  1.3E+04  &  2.0E+02   &  6.2E+04 \\
 18 56 03.87 & +02 23 23.0 & G35.586+0.061  &  mm   &  SINGLE$^\alpha$   &  1.1E+01  &  3.7E+01  &  3.5E+02  &  2.2E+03  &  4.3E+02   &  2.7E+05 \\
 18 56 05.47 & +02 22 27.0 & G35.575+0.048  &  mm   &  SED$^\alpha$   &  5.4E+00  &  2.5E+01  &  7.1E+02  &  2.0E+04  &  5.5E+01   &  1.6E+04 \\
 18 57 09.00 & +01 38 57.0 & G35.05-0.52  &  r   &  SED$^\alpha$   &  8.4E+00  &  3.6E+01  &  6.1E+02  &  7.6E+03  &  1.8E+02   &  7.0E+04 \\
 19 00 06.91 & +03 59 39.0 & G37.475-0.106  &  m   &  SINGLE$^\alpha$   &  1.0E+01  &  4.8E+01  &  8.7E+01  &  8.1E+02  &  1.6E+02   &  4.0E+05 \\
 19 00 16.00 & +04 03 07.0 & G37.55-0.11  &  r   &  SED   &  1.7E+01  &  3.9E+01  &  5.3E+02  &  1.3E+04  &  1.6E+03   &  2.3E+05 \\
 19 43 10.03 & +23 44 59.0 & G59.794+0.076$^\dag$  &  mm   &  SINGLE$^\alpha$   &  9.1E+00  &  3.4E+01  &  3.8E+01  &  3.1E+02  &  2.5E+01   &  6.5E+04 \\
 19 43 10.62 & +23 44 03.0 & G59.78+0.06$^\dag$  &  r   &  SED$^\alpha$   &  7.8E+00  &  2.6E+01  &  3.1E+02  &  2.8E+03  &  4.9E+01   &  6.8E+03 \\
\hline
\end{tabular}
\tiny
\begin{flushleft}
$^{a}$ Distance values are reported in \citet{hill05}. A $^\dag$ in this column indicates those sources which have a distance ambiguity.

$^b$ Source names given to two (or less) decimal places are consistent with those reported by \citet{walsh98, minier01, thompson06} which were targeted in the SIMBA survey (Paper I). Note that in a few instances, there may not be a direct translation between their Galactic name and their equatorial coordinates. Source names given to three decimal places, denote those sources identified by SIMBA, with the extended Galactic names intended to distinguish closely associated sources. The source names are consistent with Table 5, Paper I.

$^c$ Denotes (any) association with methanol maser sites (m) and/or \uchii\, regions (r). Those sources associated with both a methanol maser and a radio continuum source are depicted by (mr). The MM-only sources which are devoid of both of these sources are denoted (mm).\\

$^d$ This column depicts the type of SED applied to the source. `SED' indicates that a two-component spectral energy distribution has been applied, whilst `SINGLE' indicates that a single cold-component fit has been applied to the source. An $^\alpha$ in this column denotes those sources that were fit with \IRAS\, upper limits.
\end{flushleft}
\end{table*}

%

\section{Analysis}\label{sec:analysis}

In this section we discuss the method of our fitting approach and analyse each of the three parameters derived from the SED analysis: temperature, mass and luminosity with respect to the different classes of source in the sample. 

\subsection{Method}\label{sec:method}

 Spectral energy distribution diagrams have been drawn for 227 of the 405 sources appearing in the SIMBA list of Paper I (56 per cent). Of these, 135 are two-component SEDs modelling a hot and a cold component for a source, whilst the remaining 92 are single component fits which model the cold component of a source only. 

For 47 two-component SEDs, the cold component of the curve is ill-constrained as a consequence of the poor data sampling: only the 1.2\,mm SIMBA flux with \IRAS\, upper limits. In this instance the number of free parameters (T$_\mathrm{cold}$ \& M$_\mathrm{cold}$)  for the cold component exceeds the number of data points available for analysis and it is not possible to define a $\chi^2$ value nor constrain the resultant parameters from these fits. These 47 `bad' SEDs are consequently excluded from the following analysis and discussion, bringing the total number of SEDs analysed to 180 (44 per cent of the sample) and the number of two-component fits to 88.

\rmchr{ Using the Bayesian inference method for our SED fits, means that these `bad' fits should have a flat probability curve which will in turn have negligible effect on the probability curve of the entire sample. In order to test this, we excluded these 47 sources from our sample and re-produced the cumulative distribution plots for each parameter for the remaining source sample. The results indicate, as expected, that these 47 fits do not influence or affect the outcome of cumulative distribution plots for each parameter of the sample.
}

The three main parameters that we extract from the SED analysis are the dust temperature, dust mass and the luminosity. The 180 millimetre continuum sources comprise four distinct classes of source, as discussed in section \ref{sec:sample}. Table~\ref{tab:fittype} indicates how many of each class of source satisfy each of the single and two-component SEDs, as well as how many sources have a direct \IRAS\, association. Those sources that were fitted with \IRAS\, upper limits are identified by an $^\alpha$ in Table \ref{tab:sedparams}.

\begin{table}
  \begin{center}
    \caption[]{Table indicating how many of each class of source
      satisfy the different types of SEDs applied to the
      sample. Column 1 indicates whether a two-component or
      single-component SED has been applied. Column 2 indicates
      the class of the millimetre source. Column 3 indicates the
      number of sources with an \IRAS\, association, whilst column 4
      indicates the number of sources that have been fit with the
      \IRAS\, upper limits. \label{tab:fittype} }
    \begin{tabular}{@{}lclll@{}}
      \hline 
       &  \multicolumn{2}{c}{Source} &  \IRAS\,  & Upper \\
        &   \multicolumn{2}{c}{Class} &  Fit & Limit Fit \\                 
      (1) & \multicolumn{2}{c}{(2)} & (3) & (4)\\
      \hline
       TWO- & (MM) &  MM-only   &   3    & 11\\ 
           COMPONENT     & (M)&  maser     & 16      & 15\\
            (88)     & (MR)&  maser+radio &   4  & 16\\
                & (R)&  radio     &    7    & 16  \\
      \hline
      SINGLE (92) & (MM)&  MM-only & 5 & 56 \\
                  & (M)&  maser   &  1 & 18\\
                  & (MR)&  maser+radio & 1  & 8\\ 
                  & (R)&  radio   & 0 &  3\\ 
      \hline
    \end{tabular}
  \end{center}
\end{table}

 For each of the sources with a near-far distance ambiguity, the near-distance value of each parameter is assumed in the following analysis (and for comparison with Paper I). As a check, in Paper I, we examined the results for sources with no distance ambiguity with the results from the sample, assuming the near distance for 197 sources with a distance ambiguity. The results were consistent with each other and indicated the small influence of assuming the near distance value for those with an ambiguity. We thus refrain from presenting the far distance determinations of each of the parameters in this work and refer the reader to Paper~I for the far distances, and Equation \ref{eq:sed_fit} for the appropriate scaling factor where the far distance value is more applicable. Sources with a distance ambiguity are denoted by a $^\dag$ in Table \ref{tab:sedparams}.

Table \ref{tab:sedparams} presents the temperature, mass and luminosity of each of the 180 sources for which SED models were drawn. Figure \ref{fig:fit_sed} provides an example SED and probability curves for both the temperature and mass of three sources.  The top and middle panels of this figure display the respective plots for a two component SED, whilst the bottom panel illustrates a single component fit. The top panel presents a well constrained source with sharply peaked probability curves for both the temperature and mass. The middle and bottom panels on the other hand present sources with less well constrained data, i.e. the SED has had the \IRAS\, upper limits incorporated. Figure~\ref{fig:fit_sed} illustrates the need for observations spanning a broad range of the wavelength parameter space of the SED, as well as ample and tightly constrained data for obtaining sharp probability curves, \emph{i.e.} strong constraints on the parameters (for instance  G\,12.02\,-\,0.03, top panel). When less data are available (for instance, G\,0.21\,-\,0.00 and G\,10.288\,-\,0.127, middle and bottom panels, respectively) for SED analysis, the resultant probability curves are wider and uncertainties on the derived parameters are larger. In this instance, estimating the model parameters from the best fit values is simply not accurate enough.

Class-comparative histogram and cumulative distribution plots for the temperature, mass and luminosity of the sample are usually built by sorting the individual sources into bins according to the values for parameters obtained for the best SED models. However, we have shown that these ``best values'' can be attached to significant uncertainties that should be taken into account when doing statistical studies of the complete sample.

For a given source, instead of incrementing the parameter bins corresponding to the ``best value'' by 1, we increment all bins by the Bayesian probability that the parameter takes this value, with the sum of all probabilities being 1. For well constrained sources, the probability is sharp and peaks at the ``best value'' (see. Fig. \ref{fig:fit_sed}). In this instance our procedure is almost equivalent to the classical one. For less well constrained sources however, a larger number of bins are incremented by a small amount. With this method, uncertainties on the derived parameters of individual sources are automatically taken into account in the cumulative distribution plots and more weight is given to well constrained sources. That is, less well constrained sources essentially have flat distributions and they do not contribute to the shape of the histogram or cumulative plots.

\begin{figure*}
  \includegraphics[width=\hsize]{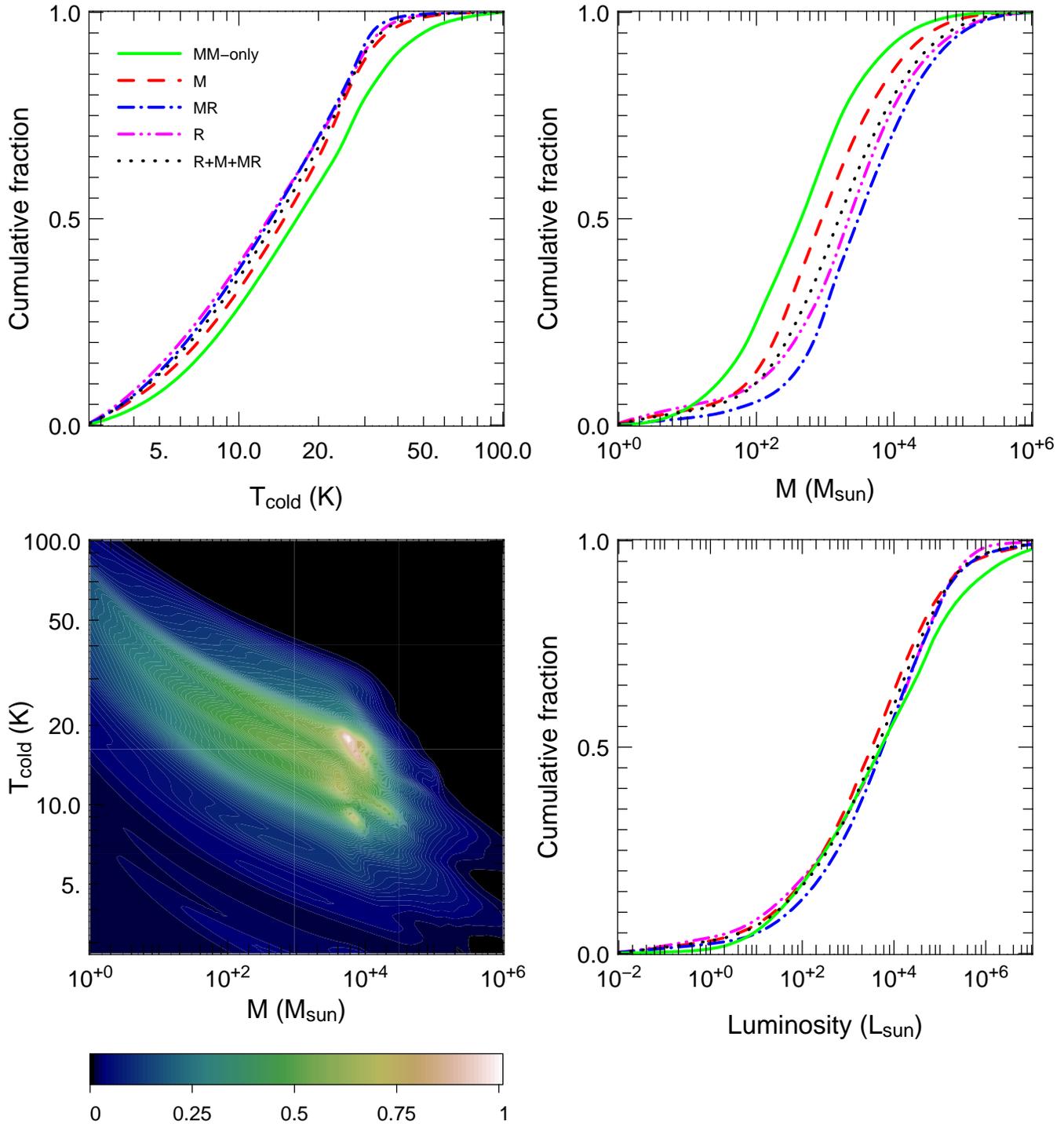}
\caption{Cumulative distributions of the SIMBA sources. The
  distributions of the individual classes are displayed on each of the plots, with the key on the top-left plot indicating which class of source is represented by which distribution.
    {\it top left }: distribution of temperature. {\it top
    right}: distribution of mass. {\it bottom left}: 2-dimensional
  probability distribution of the temperature and the mass parameters. {\it bottom
    right}: distribution of luminosity.\label{fig:cumul}}
\end{figure*}

Cumulative distribution plots for each parameter are presented in Fig. \ref{fig:cumul}. For comparative purposes, these distributions depict each of the four classes of source (see section \ref{sec:sample}) on a single plot for each parameter. 
In an attempt to ascertain whether the MM-only cores have characteristics similar to sources with star formation activity, we also drew cumulative distributions comparing the MM-only sample with the star formation activity sample (the combination of classes M, MR and R sources) for each parameter.

As we are primarily interested in the dust properties, temperature comparisons amongst the sources are made using the T$_\mathrm{cold}$ component of the sources, rather than the T$_\mathrm{hot}$ value which has only been determined for those sources with a full two-component SED. The mean and median values of the temperature, mass and luminosity for each of the classes of source, as well as the combined maser and radio continuum sources (class M, MR and R) and the entire sample are presented in Table \ref{tab:mean}.

\begin{table*}
\caption{Mean and median as derived from the SED analysis of the four
  classes of source.  For a breakdown of the number of sources within
  each class for which different SEDs were applied, refer to Table
  \ref{tab:fittype} \label{tab:mean}. }
\centering
\begin{tabular}{@{}llcccccc@{}}
\hline 
& &\multicolumn{4}{c}{Class} &whole &all except MM-only \\
Parameter& & MM-only(MM) & maser(M) & maser+radio(MR) & radio(R) & sample & (M+MR+R)\\
\hline
  temperature (K) & mean & 16.0  &  13.8  &  12.5  &  12.4  &  14.2  & 13.1 \\
               & median & 16.6  &  14.9  &  13.2  &  13.0  &  15.0  & 13.9 \\
 \hline
 mass (M$_\odot$) & mean & 4.6E+02  &  9.6E+02  &  3.3E+03  &  2.0E+03  &  9.5E+02  & 1.6E+03 \\
               & median & 4.5E+02  &  8.9E+02  &  3.0E+03  &  2.2E+03  &  9.4E+02  & 1.6E+03 \\
 \hline
 luminosity (L$_\odot$) & mean & 5.6E+03  &  2.9E+03  &  4.8E+03  &  3.3E+03  &  4.2E+03  & 3.4E+03 \\
               & median & 5.2E+03  &  3.3E+03  &  5.7E+03  &  5.4E+03
               &  4.7E+03  & 4.4E+03 \\
\hline
\end{tabular}
\end{table*}

In order to ascertain the influence of the different types of fit - single or two-component - on the cumulative distributions, we separated these two populations and drew cumulative distributions for each parameter for each fit type. The cumulative plots of each component (i.e. fit type) displayed little difference from the composite data presented in Fig. \ref{fig:cumul} for each of the temperature, mass and luminosity, and are thus not presented here.

   Kolmogorov-Smirnov (KS) tests were also performed from the cumulative distributions, in order to test the hypothesis that the MM-only sources are drawn from the same parent distribution as the other three classes of source (M, MR and R). KS tests were applied to the temperature, mass and luminosity. The results from the KS tests can be found in Table \ref{tab:kstest}, with discussion of each parameter in the following subsections.

   Generally, in order to conclude that two distributions are not drawn from the same sample, the KS probabilities must be small,  $\le$\,0.01.   As a calibration measure, the KS tests were applied to class M and R sources for each of the parameters mentioned above. The results indicate that the likelihood of these two classes of source being from different distributions, for all three parameters tested, is small (see Table \ref{tab:kstest}).

\begin{table}
\begin{center}
\caption[]{Results from the KS test of temperature, mass and luminosity for the four classes of source as well as the star formation activity sample. Column 1 indicates the parameter being tested. Columns 2 and 3 list the classes of source being tested. Column 4 gives the resultant KS probability that the objects in columns 2 and 3 are from the same parent distribution. If this  probability is $<$0.01 it is generally concluded that the samples are not drawn from the same population.  \label{tab:kstest}}
\begin{tabular}{@{}llll@{}}
\hline
Correlation & Source Class &  vs Source Class & KS-prob \\
(1) & (2) & (3) & (4) \\
\hline
Temperature & (MM) MM-only & (M) masers  & 9.2E-01 \\
&&  (MR) maser+radio & 8.5E-01 \\
&&  (R) radio  & 9.4E-01 \\
&& M+MR+R & 6.3E-01\\
& (M) maser  & (R) radio & 1.0E+00\\          
\hline
Mass & (MM) MM-only & (M) masers & 5.5E-01\\
&&  (MR) maser+radio & 4.2E-03\\ 
&&  (R) radio  & 3.9E-02\\  
&& M+MR+R & 8.6E-03 \\
& (M) maser  & (R) radio & 6.0E-01\\ 
\hline
Luminosity & (MM) MM-only & (M) maser  & 9.2E-01\\
&&  (MR) maser+radio & 1.0E+00\\
&&  (R) radio & 1.0E+00\\ 
&& M+MR+R & 9.7E-01\\ 
& (M) maser  & (R) radio & 1.0E+00 \\          
\hline
\end{tabular}
\end{center}
\end{table}

\subsection{Temperature}

The cumulative distribution plot of the temperature (Fig. \ref{fig:cumul}, {\it top, left}) indicates that there is little distinction between the four classes of source in the sample. The sources with radio continuum associations (classes MR and R) appear to be the coolest of the sample, whilst conversely the MM-only cores are the warmest of the sample. 
 We caution however that this distinction is very slight, as indicated by the KS tests which confirm that these data do not allow us to strongly discriminate between the different classes of source with respect to their temperature.

The median temperature of the sample (i.e. the value corresponding to the 0.5 fraction value on Fig. \ref{fig:cumul}) of the MM-only sample is 17\,K, the maser sample (class M) is 15\,K,  the maser+radio sample (class MR) is 13\,K, and the radio sample (class R) is 13\,K.
 These median values are not inconsistent with the 20\,K temperature assumed in Paper~I for purposes of mass derivation and they emphasize the small difference between the samples in terms of temperature.  The cumulative distribution of the sources with known star formation activity (classes M, MR and R) indicates that collectively these sources are marginally cooler on average than class MM-only sources i.e. the sources apparently without star formation activity.

The shape of each of the individual class distributions in a cumulative plot is also a useful diagnostic tool. The linear shape of the temperature profile of classes MR and R sources indicates that there is little constraint on their temperature as determined from our SED analysis. Only an upper limit of 40\,-\,50\,K can be firmly established. We also draw attention to the fact that the cumulative distribution of the temperature shows values $<$5\,K. As the cumulative distribution reports the sum of all probabilities over all temperatures, this is simply the tail end of the temperature distribution and we do not attribute any significance to these low temperatures.

Interestingly, the cumulative distribution of the maser sample (class M) directly traces the temperature profile of the entire sample (not shown). This perhaps indicates that the maser sample displays a global temperature profile typical of a large cross section of massive star formation sources, and more specifically our entire sample. That is, the maser population may be considered the `standard' massive star formation population (this is explored further in section \ref{sec:mvl}).

\subsection{Mass}

   The cumulative distribution plot of the mass (Fig. \ref{fig:cumul}, {\it top, right}) indicates that MM-only sources are the least massive of the sample, whilst the sources with both a methanol maser and radio continuum source (class MR) are the most massive of the sample. As per the temperature distribution, the maser sample again traces the mass distribution of the whole sample (not shown).

From these cumulative distributions and the KS-tests (Table \ref{tab:kstest}), the null hypothesis that the samples are drawn from the same population can clearly be rejected when comparing the MM-only sample with those sources associated with both a methanol maser and radio continuum source (i.e. class MR). This suggests that the MM-only sample is not from the same parent distribution as sources with both a maser and radio continuum source for the mass parameter. 
Notably, the null hypothesis can also be rejected when comparing the MM-only sources with the combined star formation sample (class M + MR + R). This is not surprising considering that this sample is comprised of sources that have already been proven to be distinct from the MM-only sample (i.e. class MR). There is also a weak suggestion that the MM-only sample is not from the same distribution as sources with a radio continuum association (class R) for the mass. No distinctions could be discerned regarding the mass of the methanol maser sample (class M).

   If we compare the cumulative mass distribution (Fig. \ref{fig:cumul}, middle) with the same plot produced in Paper I (Fig. 4) for an assumed temperature of 20\,K, we find many similarities. Fig. 4 of Paper I showed that the MM-only sources were the least massive of the sample, followed by the methanol maser sources, the radio continuum sources, with the class MR sources the most massive in the sample. This result also holds true for the cumulative mass plot in this paper, which depicts a more robust estimate of the mass, taking into account the uncertainties on the temperature, rather than assuming a fixed one.

  The cumulative distribution of the combined star formation activity sources (classes M, MR and R) indicates that these sources are more massive on average than the MM-only sources. Median values for each sample are given in Table \ref{tab:mean} which further corroborates this result.

   Figure \ref{fig:cumul} ({\it bottom, left}) shows the 2-D probability plot of the temperature and the mass. It is clear from this plot that these two parameters are highly correlated, which is hardly surprising given the nature of the modelling. 
 The shading on the plot indicates the probability of occurrence for both the temperature and mass of the source in our sample, marginalizing the contribution of the hot component for those sources where it is applicable.
 From this plot it is possible to ascertain the most likely combination of temperature and mass values for the sources in our sample. 

The 2-D probability plot corroborates the cumulative plot of temperature  indicating the low probability of the sources in our sample having temperatures of $<$\,5\,K.  The most likely temperature of the sources in our sample is between 10 and 20\,K, again confirming that our sources are cold, as well as the results of our earlier work in Paper I.  The 2-D probability plot also indicates that the sources in our sample span a wide range of mass values, with the most probable mass roughly around  10$^4$\,\mstar. The most likely combination of these parameters for our sources is a temperature of 17\,K and a mass of 5.7\,$\times$\,10$^4$\,\mstar. This plot also illustrates that the very low and high temperatures ($\lesssim$\,5\,K and $\gtrsim$\,50\,K respectively) correspond to very high and low masses ($\gtrsim$\,10$^5$\,\mstar\, and $\lesssim$ 10\,\mstar, respectively). Although, these combinations of parameters are plausible solutions to the SED analysis, they do not correspond to physically meaningful solutions, and the low, but non negligible, associated probabilities should be interpreted with care. 

\subsection{Luminosity}

   The cumulative distribution of the source luminosity is also presented in Fig. \ref{fig:cumul} ({\it bottom, right}). These plots, in addition to the KS tests, indicate that there is little difference, if any, between the different classes of sources in the sample in terms of their luminosity. Again, we attribute little significance to low luminosity values in the distribution, which are simply the tail-end values of the probabilities which accordingly have a low probability of occurrence. Interestingly, the distributions for each of the different classes of source in the sample do not display the same shape or gradient. 
 The mean and median values of the luminosity for each of the different classes in the sample are presented in Table \ref{tab:mean}.


\section{Discussion}

\subsection{Massive star formation: luminosity, mass and SED modelling}

Identification and characterisation of young massive stars at all stages of their evolution, especially of the earliest stages, is essential in addressing and defining evolutionary scenarios for their formation. Among the meaningful physical parameters, the mass and luminosity estimates of massive star-forming cores could play a pivotal role in their characterisation and classification, possibly providing insight into their evolutionary status in a similar way that they do for low-mass protostar classification from class 0 to class I objects \citep[e.g.][]{andre00}.

There are many different data analysis techniques to determine the mass and luminosity of candidate young massive (proto)stars. Assuming a temperature, the mass estimate is usually derived from knowledge of the (sub)millimetre continuum fluxes (e.g. \citealp{motte01, hill05}). Whilst assuming a temperature across a source sample is useful for first order approximations of the physical properties (e.g. mass) of a source, in reality all sources in a region \citep{motte01} or survey \citep{hill05} will not be at the same global temperature. We cautioned in Paper I that temperature assumption for mass determinations could lead to under- or over-estimations of the mass if the temperature was in fact cooler or warmer, respectively. It should be expected that different classes of source, such as methanol maser and radio continuum sources, which originate under different conditions, will also have different physical characteristics --  including (but not limited to) their temperature and mass. For a large sample size, comprised of different classes of source, we therefore should not expect a global temperature to apply to a particular star formation complex or sample.

SED modelling could, in contrast, provide us with useful estimates of the luminosity, temperature, and mass of star-forming cores if the observational data are well constrained (e.g. \citealp{burton04, minier05}). In this instance, data are typically fitted with SED components that reproduce the cold core emission in the far-infrared-millimetre domain and the  warm infrared emission that could be caused by a cluster of young stars. The shape of a SED, as well as the position of the SED maxima in the far-infrared-submillimetre and mid-infrared domains, can be indicative of evolutionary stage \citep[e.g.][]{minier05}. Compiling a high quality well-constrained data set for spectral energy distribution modelling \citep[cf.][]{minier05} is, however, difficult, especially with the present lack of suitable high-resolution data covering the peak of the SED at $\sim$\,100\,\microm. The extension of this method to large source numbers makes this an arduous task at best. This method is limited to small numbers of well-studied individual sources.

Various radiative transfer models have also been developed which build large grids of SED models for young stellar objects (YSOs) to which observational data can be compared and fitted. However, radiative transfer models, such as those of \citet{whitney03} and \citet{robitaille06}, have been mainly developed for low mass (proto)stars. These authors are principally concerned with modelling mid-infrared emission from low-mass young stellar objects in relatively nearby star forming regions. Their conclusions rest on the characteristics of the warm, mid-infrared dust emission surrounding these (proto)stars, and their orientations to our line of sight. Applying these SED fitting techniques has recently been attempted by \citet{molinari08}. However, there are many caveats and limitations to SED modelling of young massive (proto)stars such as the presence of multiple sources and the confusion due to large distances \citep{robitaille08}, which need to be taken into consideration.

\subsection{Our approach: SED modelling with Bayesian inference method}

In our approach to SED analysis, we adopted a very simple description for the calculations of synthetic SEDs, rather than using detailed radiative transfer modelling. This approach was chosen for two reasons:

Detailed radiative transfer modelling requires a large number of free parameters (generally more than 10) which describe the spatial distribution of the dust grains and the source illumination, for example. When the available observational data is limited, this kind of modelling results in degeneracies between each of these parameters, preventing firm constraints from being established. As we have only partial wavelength coverage of the spectral energy distribution, only the characteristic parameters of our sources, such as the temperature, mass and luminosity, can be extracted in a quantitative way. We thus chose to restrict the number of free parameters for our SED modelling. Although providing a less accurate description, this simple model provides a valid alternative to estimate the main parameters of the sources. 

In addition, only simple, analytical models allow systematic sampling of the parameter space in a reasonable period of time (6.25\,$\times$\,10$^6$ models in a four-dimensional parameter space). This subsequently allows statistical studies of a large number of models, via Bayesian inference, to be undertaken in order to determine the interplay between parameters and establish robust ranges of validity on the parameters. Such a work, especially when dealing with a large number of sources, is beyond current modelling capacities if radiative transfer models are used. For instance, a large numerical effort of SED modelling of YSOs has been undertaken by \cite{robitaille06}, whose current grid of models includes a total number of 200\,000 SED models, sampling a 14-dimensional parameter space. This limited number of models cannot provide a systematic sampling of the parameter space and the authors needed to bias the explored parameter space to accommodate the computing time required. As a result,  trends and correlations between parameters are difficult, if not impossible, to apprehend with such a grid of models (see \cite{robitaille08} for a more detailed description of the limitations in the use of their grid of models to estimate parameters).

\subsection{Interpreting the distinction between mass and luminosity of our source samples}\label{sec:mvl}

Using the first application of the Bayesian inference method of SED modelling to massive star formation,  we have estimated a range of values for the temperature, mass and luminosity for sources from the SIMBA survey of \citet{hill05}. We have performed a statistically robust analysis of each of these parameters across the different classes of source in the sample, with the aim of characterising and classifying them.

   If we attribute some significance to the small distinction between the cumulative distributions of the various object classes, MM-only cores to \uchii\, regions (Fig. \ref{fig:cumul} and Table \ref{tab:kstest}), two conclusions can be reached. Firstly, the cumulative distribution of the methanol maser sources, for each parameter, is very much representative of the entire sample.  This result (Fig. \ref{fig:cumul}) coupled with the fact that methanol masers are exclusive signatures of high mass star formation confirms that our sample is representative of high mass star formation. In addition, the MM-only cores are less massive on average than cores with a methanol maser and/or radio continuum association and hence those sources known to support massive star formation. This is in agreement with Paper I, despite the different approaches to mass determination. The cumulative distribution plots of the temperature and luminosity (Fig. \ref{fig:cumul}) display only small distinctions between the different classes of source in the sample, which is more the case for the luminosity than the temperature. Notably, there is no significant difference between the MM-only sample and the combined star formation sources, i.e those sources with a methanol maser and/or radio continuum source (class M, MR and R). This result suggests that the MM-only sample has similar characteristics as sources with known star formation sites. However, the KS-tests indicate that the MM-only cores are not from the same parent population as sources with both a methanol maser and radio continuum association (class MR) with respect to their masses.

\begin{figure}
  \includegraphics[width=\hsize]{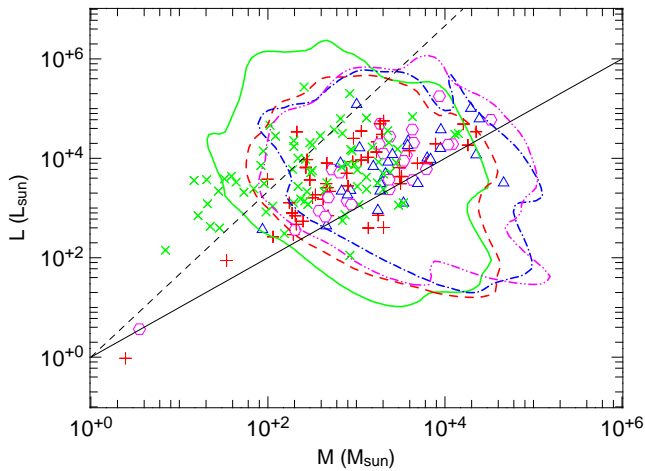}
  \caption{Mass-Luminosity diagram for the different classes of source in the sample. Each of the points represents the median value  (\emph{i.e.} the values at which the cumulative probability distribution is 0.5) of the mass and luminosity of the individual sources (green cross: mm, red plus sign: m, blue triangles:mr and pink circles: r).  The contour levels represent the regions enclosing 68 per cent of the Bayesian probability for a given class of sources (green full line: mm, red dashed line: m, blue dot-dash line: mr and pink dot-dot-dash line: r). The full and dashed straight lines depict the $M = L$ and $M = L^{0.6}$ relations, respectively.\label{fig:L_M}}
\end{figure}

Figure \ref{fig:L_M} introduces a luminosity vs. mass diagram (hereafter, $M-L$ diagram) for our sample of cores. A similar diagram was proposed for class 0 and class I low-mass protostars by \citet{andre00}, who presented the protostellar envelope mass vs. the bolometric luminosity for individual protostars or multiple protostar systems. Comparatively, in this work, the derived masses in Table~\ref{tab:sedparams} overestimate any protostellar envelope mass because the (sub)millimetre emission fluxes are integrated over more than a FWHM beam. The masses reported in Table \ref{tab:sedparams}  almost certainly include contributions from both the core and from diffuse extended gas, whilst the luminosities are rough estimates of the source bolometric luminosity.
  Consequently, if a dominant source is responsible for both the luminosity and the mass of the core, then the $M-L$ diagram could prove a useful diagnostic of the evolutionary status of this dominant object. 

Figure \ref{fig:L_M} plots the median value (the value corresponding to 0.5 on the cumulative distributions) of mass and luminosity for each of the sources in the sample, as well as the 68 per cent Bayesian probability of occurrence for each source class (the latter is explored in the following section).
 The masses and luminosities span $10\,-\,10^4$\,\mstar\, and $10^2\,-\,10^6$\lum, respectively. These ranges are comparable to those presented in Fig. 9 of \citet{molinari08}, although our objects have higher masses on average. \citet{molinari08} undertook a method of approach different to our own, in which they assigned millimetre emission fluxes to the massive star-forming objects by removing emission from the diffuse clumpy medium. Their resultant masses are consequently lower for their sources compared with this work.  They also fitted their SED to a grid of models that were computed according to \citet{whitney03}.

  In Fig. \ref{fig:L_M} the MM-only cores appear to segregate into two groups with the division loosely defined by the line of M\,=\,L$^{0.6}$. The first group of MM-only sources lies in the same region of the plot as the sources with known star formation activity (methanol maser and/or radio continuum sources) i.e. classes M, MR and R. The other group of MM-only sources exhibit lower masses or higher luminosities and are clearly separated from class M, MR and R sources. 

 In summary, the cumulative distribution plots produced from our SED models as well as the KS-tests and the $M-L$ diagram reveal that some MM-only sources are distinct from those sources with a radio continuum association in terms of the mass parameter. These results are consistent with Paper I. In Paper I, it was also revealed that the MM-only cores have the smallest radii of the sample and at least 45 per cent are without mid-infrared \MSX\, emission. The characteristics of some MM-only cores $-$ the least luminous and infrared quiet ones $-$ are consistent with less evolved examples of massive star formation. However, we cannot exclude the possibility that some MM-only cores may be dominated by luminous intermediate-mass protostars, hence less massive but still relatively luminous cores. Finally, some fraction of the MM-only cores may even represent the `failed' cores hypothesised by \citet{vaz05}. More likely, the MM-only cores represent various classes of prestellar and protostellar sources.

\subsection{Significance of luminosity vs. mass diagram}

Alternatively, if we do not attribute any significance to the small distinction between the various source samples in Fig.~\ref{fig:cumul}, these results may simply reflect the limited information that can be extracted from our SED analysis, especially considering the  poorly constrained data in the far-infrared regime. There are many limitations to the interpretations of spectral energy distributions and $M-L$ diagrams for high-mass star forming cores. For instance, the luminosity estimates are poorly constrained in the far-infrared domain and the angular resolution of SEST at 1.2\,mm is insufficient to exclude the possibility of multiple source components - indeed \citet{longmore06} have detected a few massive infrared protostars or young stellar objects within 6,000\,AU, corresponding to 24 arcsec at 4\,kpc (i.e. with our current data, we would not expect to see this).

To account for this, Figure \ref{fig:L_M} also presents the 2-D Bayesian probability contours enclosing the 68 per cent Bayesian probability of occurrence for both the mass and the luminosity. It can be seen that the mass and the luminosity encompass a much larger range of values than depicted by the median values on the plot (see section \ref{sec:mvl}).
  The MM-only cores could occupy the range 100\,\mstar\, and $10^6$\,\lum\, or $10^4$\,\mstar\, and $10$\,\lum\,  of Fig. \ref{fig:L_M} with equal probability.
  That is, these results suggest that the MM-only cores could either be star-forming regions in quite an advanced stage hosting embedded, luminous stars ($10^6$\,\lum) or alternatively very massive quiescent clouds ($10$\,\lum\,).
 Moreover, we observe a similarly wide range of probable luminosity and mass for the \uchii\, regions and the methanol maser sources in our sample, despite the fact that these sources are well identified at high angular resolution. The Bayesian inference method demonstrates that SED fitting, with the currently available far-infrared data sets, cannot provide us with reliable evolutionary tracks in the $M-L$ diagram for high-mass star formation.

\section{Conclusions}

We have performed spectral energy distribution analysis for 227 of the 405 sources detected in the SIMBA survey of \citet{hill05}. Using the Bayesian inference method of analysis we have determined a range of suitable values, with associated probabilities, for each of the parameters of temperature, mass and luminosity. Each of these parameters have been analysed with respect to the different type of source in the sample and hence their associations (or lack thereof) with methanol maser and/or radio continuum sources. The cumulative distribution plot of the mass for the different source classes is consistent with our earlier work.

  If we attribute little significance to the class distinction for each parameter of mass, luminosity and temperature, then the MM-only cores have the same characteristics as sources with known star formation activity (classes M, MR and R), yet they display no overt signs of star formation. Following this, the MM-only cores are excellent candidates for early stage protostars or massive young stellar objects.

Attributing significance to the marginal distinctions between the MM-only sources and those sources with methanol maser and/or radio continuum associations for each of the temperature and luminosity, and factoring in the results of the mass, radius and lack of mid-infrared associations for almost half, then we can interpret that the MM-only core is a younger, smaller and less evolved example of massive star formation. That is, they represent an evolutionary stage of massive star formation, prior to the development of methanol maser emission and are thus indicative of the earliest stages of massive star evolution.  However, we cannot exclude the possibility that the MM-only cores are examples of `failed' cores or instead will support intermediate mass star formation. Alternatively, the MM-only core could comprise a cross-section of sources supporting both arguments. Spectral line observations (e.g. of turbulent linewidths and/or chemical state) of these MM-only cores are necessary in order to determine which of them, if any are forming massive stars.

It is clear from this work, that SED modelling is heavily reliant upon well-constrained and robust data which is well-sampled in wavelength space. There is a clear dependency of the stringency of the fit upon the quality of the data. From our Bayesian inference SED analysis, it is clear that in the absence of reliable far-infrared data, which would serve to constrain the peak of the SED, it is not possible to draw reliable evolutionary tracks in the mass versus luminosity diagram of high-mass star formation. Future observations with the Herschel Space Observatory will provide greater constraints in the crucial far-infrared/submillimetre regimes for\ SED modelling.

\section*{Acknowledgments}
  We would like to thank S. Lumsden for his \MSX\, script, as well as S. Longmore and C. Purcell for assistance with converting the \MSX\, and \IRAS\, images. We wish to thank P. Jones and H. J. van Langevelde for useful discussions regarding the SED fits. We extend thanks to an anonymous referee for their useful improvements to the manuscript. C.~Pinte acknowledges funding support of the European Commission's Seventh Framework Program as a Marie Curie Intra-European Fellow (PIEF-GA-2008-220891).  This research has made use of the NASA/ IPAC Infrared Science Archive, which is operated by the Jet Propulsion Laboratory, California Institute of Technology, under contract with the National Aeronautics and Space Administration. This work has made use of the image production toolkit {\sc karma}.

\bibliographystyle{mn2e}
\bibliography{/Users/thill/data/bib/aa,/Users/thill/data/bib/references}

\begin{thebibliography}{}

\bibitem[\protect\citeauthoryear{{Andr\'e}, {Ward-Thompson} \&
  {Barsony}}{{Andr\'e} et~al.}{2000}]{andre00}
{Andr\'e} P.,  {Ward-Thompson} D.,    {Barsony} M.,  2000, Protostars and
  Planets IV, p.~59

\bibitem[\protect\citeauthoryear{{Batrla}, {Matthews}, {Menten} \&
  {Walmsley}}{{Batrla} et~al.}{1987}]{batrla87}
{Batrla} W.,  {Matthews} H.~E.,  {Menten} K.~M.,    {Walmsley} C.~M.,  1987,
  Nature, 326, 49

\bibitem[\protect\citeauthoryear{{Beichman}, {Neugebauer}, {Habing}, {Clegg} \&
  {Chester}}{{Beichman} et~al.}{1988}]{beichman88}
{Beichman} C.~A.,  {Neugebauer} G.,  {Habing} H.~J.,  {Clegg} P.~E.,
  {Chester} T.~J.,  eds, 1988, {Infrared astronomical satellite (IRAS) catalogs
  and atlases. Volume 1: Explanatory supplement}

\bibitem[\protect\citeauthoryear{{Beuther}, {Schilke}, {Menten}, {Motte},
  {Sridharan} \& {Wyrowski}}{{Beuther} et~al.}{2002}]{beuther02}
{Beuther} H.,  {Schilke} P.,  {Menten} K.~M.,  {Motte} F.,  {Sridharan} T.~K.,
    {Wyrowski} F.,  2002, ApJ, 566, 945

\bibitem[\protect\citeauthoryear{{Burton}, {Lazendic}, {Yusef-Zadeh} \&
  {Wardle}}{{Burton} et~al.}{2004}]{burton04}
{Burton} M.~G.,  {Lazendic} J.~S.,  {Yusef-Zadeh} F.,    {Wardle} M.,  2004,
  MNRAS, 348, 638

\bibitem[\protect\citeauthoryear{{Caswell}, {Vaile}, {Ellingsen}, {Whiteoak} \&
  {Norris}}{{Caswell} et~al.}{1995}]{caswell95}
{Caswell} J.~L.,  {Vaile} R.~A.,  {Ellingsen} S.~P.,  {Whiteoak} J.~B.,
  {Norris} R.~P.,  1995, MNRAS, 272, 96

\bibitem[\protect\citeauthoryear{{Fa{\' u}ndez}, {Bronfman}, {Garay}, {Chini},
  {Nyman} \& {May}}{{Fa{\' u}ndez} et~al.}{2004}]{faundez04}
{Fa{\' u}ndez} S.,  {Bronfman} L.,  {Garay} G.,  {Chini} R.,  {Nyman} L.~A.,
  {May} J.,  2004, A\&A, 426, 97

\bibitem[\protect\citeauthoryear{{Hill}, {Burton}, {Minier}, {Thompson},
  {Walsh}, {Hunt-Cunningham} \& {Garay}}{{Hill} et~al.}{2005}]{hill05}
{Hill} T.,  {Burton} M.~G.,  {Minier} V.,  {Thompson} M.~A.,  {Walsh} A.~J.,
  {Hunt-Cunningham} M.,    {Garay} G.,  2005, MNRAS, 363, 405

\bibitem[\protect\citeauthoryear{{Hill}, {Thompson}, {Burton}, {Walsh},
  {Minier}, {Cunningham} \& {Pierce-Price}}{{Hill} et~al.}{2006}]{hill06}
{Hill} T.,  {Thompson} M.~A.,  {Burton} M.~G.,  {Walsh} A.~J.,  {Minier} V.,
  {Cunningham} M.~R.,    {Pierce-Price} D.,  2006, MNRAS, 368, 1223

\bibitem[\protect\citeauthoryear{{Lay}, {Carlstrom} \& {Hills}}{{Lay}
  et~al.}{1997}]{lay97}
{Lay} O.~P.,  {Carlstrom} J.~E.,    {Hills} R.~E.,  1997, ApJ, 489, 917

\bibitem[\protect\citeauthoryear{{Longmore}, {Burton}, {Barnes}, {Wong},
  {Purcell} \& {Ott}}{{Longmore} et~al.}{2007}]{longmore07}
{Longmore} S.~N.,  {Burton} M.~G.,  {Barnes} P.~J.,  {Wong} T.,  {Purcell}
  C.~R.,    {Ott} J.,  2007, MNRAS, 379, 535

\bibitem[\protect\citeauthoryear{{Longmore}, {Burton}, {Minier} \&
  {Walsh}}{{Longmore} et~al.}{2006}]{longmore06}
{Longmore} S.~N.,  {Burton} M.~G.,  {Minier} V.,    {Walsh} A.~J.,  2006,
  MNRAS, 369, 1196

\bibitem[\protect\citeauthoryear{{Lumsden}, {Hoare}, {Oudmaijer} \&
  {Richards}}{{Lumsden} et~al.}{2002}]{lumsden02}
{Lumsden} S.~L.,  {Hoare} M.~G.,  {Oudmaijer} R.~D.,    {Richards} D.,  2002,
  MNRAS, 336, 621

\bibitem[\protect\citeauthoryear{{Minier}, {Burton}, {Hill}, {Pestalozzi},
  {Purcell}, {Garay}, {Walsh} \& {Longmore}}{{Minier} et~al.}{2005}]{minier05}
{Minier} V.,  {Burton} M.~G.,  {Hill} T.,  {Pestalozzi} M.~R.,  {Purcell}
  C.~R.,  {Garay} G.,  {Walsh} A.~J.,    {Longmore} S.,  2005, A\&A, 429, 945

\bibitem[\protect\citeauthoryear{{Minier}, {Conway} \& {Booth}}{{Minier}
  et~al.}{2001}]{minier01}
{Minier} V.,  {Conway} J.~E.,    {Booth} R.~S.,  2001, A\&A, 369, 278

\bibitem[\protect\citeauthoryear{{Molinari}, {Pezzuto}, {Cesaroni}, {Brand},
  {Faustini} \& {Testi}}{{Molinari} et~al.}{2008}]{molinari08}
{Molinari} S.,  {Pezzuto} S.,  {Cesaroni} R.,  {Brand} J.,  {Faustini} F.,
  {Testi} L.,  2008, A\&A, 481, 345

\bibitem[\protect\citeauthoryear{{Motte} \& {Andr{\'e}}}{{Motte} \&
  {Andr{\'e}}}{2001}]{motte01}
{Motte} F.,  {Andr{\'e}} P.,  2001, A\&A, 365, 440

\bibitem[\protect\citeauthoryear{{Olmi}, {Cesaroni}, {Neri} \&
  {Walmsley}}{{Olmi} et~al.}{1996}]{olmi96}
{Olmi} L.,  {Cesaroni} R.,  {Neri} R.,    {Walmsley} C.~M.,  1996, A\&A, 315,
  565

\bibitem[\protect\citeauthoryear{{Osorio}, {Lizano} \& {D'Alessio}}{{Osorio}
  et~al.}{1999}]{osorio99}
{Osorio} M.,  {Lizano} S.,    {D'Alessio} P.,  1999, ApJ, 525, 808

\bibitem[\protect\citeauthoryear{{Ossenkopf} \& {Henning}}{{Ossenkopf} \&
  {Henning}}{1994}]{ossenkopf94}
{Ossenkopf} V.,  {Henning} T.,  1994, A\&A, 291, 943

\bibitem[\protect\citeauthoryear{{Pestalozzi}, {Minier} \&
  {Booth}}{{Pestalozzi} et~al.}{2005}]{pestalozzi05}
{Pestalozzi} M.~R.,  {Minier} V.,    {Booth} R.~S.,  2005, A\&A, 432, 737

\bibitem[\protect\citeauthoryear{{Pierce-Price}, {Richer}, {Greaves},
  {Holland}, {Jenness}, {Lasenby}, {White}, {Matthews}, {Ward-Thompson},
  {Dent}, {Zylka}, {Mezger}, {Hasegawa}, {Oka}, {Omont} \&
  {Gilmore}}{{Pierce-Price} et~al.}{2000}]{p-p00}
{Pierce-Price} D.,  {Richer} J.~S.,  {Greaves} J.~S.,  {Holland} W.~S.,
  {Jenness} T.,  {Lasenby} A.~N.,  {White} G.~J.,  {Matthews} H.~E.,
  {Ward-Thompson} D.,  {Dent} W.~R.~F.,  {Zylka} R.,  {Mezger} P.,  {Hasegawa}
  T.,  {Oka} T.,  {Omont} A.,    {Gilmore} G.,  2000, ApJL, 545, L121

\bibitem[\protect\citeauthoryear{{Pinte}, {Fouchet}, {M{\'e}nard}, {Gonzalez}
  \& {Duch{\^e}ne}}{{Pinte} et~al.}{2007}]{pinte07}
{Pinte} C.,  {Fouchet} L.,  {M{\'e}nard} F.,  {Gonzalez} J.-F.,
  {Duch{\^e}ne} G.,  2007, A\&A, 469, 963

\bibitem[\protect\citeauthoryear{{Pinte et al.}}{{Pinte et
  al.}}{2008}]{pinte08}
{Pinte et al.} C.,  2008, A\&A, 489, 633

\bibitem[\protect\citeauthoryear{{Press}, {Teukolsky}, {Vetterling} \&
  {Flannery}}{{Press} et~al.}{1992}]{press92}
{Press} W.~H.,  {Teukolsky} S.~A.,  {Vetterling} W.~T.,    {Flannery} B.~P.,
  1992, {Numerical recipes in C. The art of scientific computing}.
Cambridge: University Press, |c1992, 2nd ed.

\bibitem[\protect\citeauthoryear{{Price}}{{Price}}{1995}]{price95}
{Price} S.~D.,  1995, Space Science Reviews, 74, 81

\bibitem[\protect\citeauthoryear{{Price}, {Egan}, {Carey}, {Mizuno} \&
  {Kuchar}}{{Price} et~al.}{2001}]{price01}
{Price} S.~D.,  {Egan} M.~P.,  {Carey} S.~J.,  {Mizuno} D.~R.,    {Kuchar}
  T.~A.,  2001, AJ, 121, 2819

\bibitem[\protect\citeauthoryear{{Rathborne}, {Simon} \& {Jackson}}{{Rathborne}
  et~al.}{2007}]{rathborne07}
{Rathborne} J.~M.,  {Simon} R.,    {Jackson} J.~M.,  2007, ApJ, 662, 1082

\bibitem[\protect\citeauthoryear{{Robitaille}}{{Robitaille}}{2008}]{robitaille%
08}
{Robitaille} T.~P.,  2008, in {Beuther} H.,  {Linz} H.,   {Henning} T.,  eds,
  Massive Star Formation: Observations Confront Theory Vol.~387 of Astronomical
  Society of the Pacific Conference Series, {SED Modeling of Young Massive
  Stars}.
pp 290--297

\bibitem[\protect\citeauthoryear{{Robitaille}, {Whitney}, {Indebetouw}, {Wood}
  \& {Denzmore}}{{Robitaille} et~al.}{2006}]{robitaille06}
{Robitaille} T.~P.,  {Whitney} B.~A.,  {Indebetouw} R.,  {Wood} K.,
  {Denzmore} P.,  2006, Astrophys. J. Supp. Series, 167, 256

\bibitem[\protect\citeauthoryear{{Thompson}, {Hatchell}, {Walsh}, {MacDonald}
  \& {Millar}}{{Thompson} et~al.}{2006}]{thompson06}
{Thompson} M.~A.,  {Hatchell} J.,  {Walsh} A.~J.,  {MacDonald} G.~H.,
  {Millar} T.~J.,  2006, A\&A, 453, 1003

\bibitem[\protect\citeauthoryear{{V{\'a}zquez-Semadeni}, {Kim}, {Shadmehri} \&
  {Ballesteros-Paredes}}{{V{\'a}zquez-Semadeni} et~al.}{2005}]{vaz05}
{V{\'a}zquez-Semadeni} E.,  {Kim} J.,  {Shadmehri} M.,    {Ballesteros-Paredes}
  J.,  2005, ApJ, 618, 344

\bibitem[\protect\citeauthoryear{{Walsh}, {Burton}, {Hyland} \&
  {Robinson}}{{Walsh} et~al.}{1998}]{walsh98}
{Walsh} A.~J.,  {Burton} M.~G.,  {Hyland} A.~R.,    {Robinson} G.,  1998,
  MNRAS, 301, 640

\bibitem[\protect\citeauthoryear{{Walsh}, {Macdonald}, {Alvey}, {Burton} \&
  {Lee}}{{Walsh} et~al.}{2003}]{walsh03}
{Walsh} A.~J.,  {Macdonald} G.~H.,  {Alvey} N.~D.~S.,  {Burton} M.~G.,    {Lee}
  J.-K.,  2003, A\&A, 410, 597

\bibitem[\protect\citeauthoryear{{Whitney}, {Wood}, {Bjorkman} \&
  {Cohen}}{{Whitney} et~al.}{2003}]{whitney03}
{Whitney} B.~A.,  {Wood} K.,  {Bjorkman} J.~E.,    {Cohen} M.,  2003, ApJ, 598,
  1079

\bibitem[\protect\citeauthoryear{{Williams}, {Fuller} \&
  {Sridharan}}{{Williams} et~al.}{2004}]{williams04}
{Williams} S.~J.,  {Fuller} G.~A.,    {Sridharan} T.~K.,  2004, A\&A, 417, 115

\bibitem[\protect\citeauthoryear{{Wood} \& {Churchwell}}{{Wood} \&
  {Churchwell}}{1989}]{wood89morph}
{Wood} D.~O.~S.,  {Churchwell} E.,  1989, ApJS, 69, 831

\end{thebibliography}
\expandafter\ifx\csname natexlab\endcsname\relax\def\natexlab#1{#1}\fi

\appendix

\bsp

\label{lastpage}

\end{document}